\documentclass[12pt]{article}
\usepackage [cp1251]{inputenc}
\usepackage [english]{babel}
\usepackage{graphicx}
\usepackage {amsmath}
\usepackage {amsfonts}
\usepackage {amssymb}
\usepackage {mathrsfs}
\usepackage {longtable}
\title{Generation of magnetic fields by thermomagnetic effects in a nonuniformly rotating layer of an electrically conductive fluid}
\author{$^{1}$\textbf{M.I. Kopp},$^{1}$\textbf{K.N. Kulik}, $^3$\textbf{A.V. Tur}, $^{1,2}$\textbf{V.V. Yanovsky}}
\begin{document}

\maketitle

$^{1}$ \textit{Institute for Single Crystals, NAS  Ukraine, Nauky Ave. 60, Kharkov 61001, Ukraine}

$^{2}$\textit{V.N. Karazin Kharkiv National University 4 Svobody Sq., Kharkov 61022, Ukraine}

$^{3}$\textit{Universit\'{e} de Toulouse [UPS], CNRS, Institut de Recherche en Astrophysique et Plan\'{e}tologie,
9 avenue du Colonel Roche, BP 44346, 31028 Toulouse Cedex 4, France}

\abstract{In this paper, the generation of magnetic fields in a nonuniformly rotating layer of finite thickness of an electrically conducting fluid by thermomagnetic (TM) instability. This instability arises due to the temperature gradient  $\nabla T_0$ and thermoelectromotive coefficient gradient  $\nabla\alpha $.  The influence of the generation of a toroidal magnetic field by TM instability on convective instability in a nonuniformly rotating layer of an electrically conductive fluid in the presence of a vertical constant magnetic field ${\bf{B}}_0 \| {\rm OZ}$  is established.  As a result of applying the method of perturbation theory for the small parameter  $ \epsilon = \sqrt {(\textrm {Ra}-\textrm {Ra}_c) / \textrm {Ra}_c} $ of supercriticality of the stationary Rayleigh number  $\textrm {Ra}_c$  a nonlinear equation of the Ginzburg-Landau type was obtained. This equation describes the evolution of the finite amplitude of perturbations. Numerical solutions of this equation made it possible to determine the heat transfer in the fluid layer with and without TM effects. It is shown that the amplitude of the stationary toroidal magnetic field noticeably increases with allowance for TM effects.}


\section{Introduction}

Magnetic fields in the Universe are observed at various cosmological scales from small planets to galaxies. The question of the origin of magnetic fields in various astrophysical objects is dealt with by the dynamo theory, which was formed as an independent section of magnetohydrodynamics. For the first time, the term "dynamo"  was coined by Sir Larmor \cite{1s}. In his opinion, the hydrodynamic motion of an electroconductive fluid can generate a magnetic field by acting as a dynamo. In the modern sense, the dynamo mechanism represents the generation of a large-scale magnetic field due to the combined action of the differential (nonuniform) rotation of an astrophysical object and the mirror asymmetry of turbulent or convective flows. The topological characteristic of such flows is helicity $J_s=\overline{{\bf v} \textrm{rot}{\bf v}}$, which is a measure of knottedness of vortex field force lines \cite{2s}.
The generation of the mean field occurs under the action of a turbulent e.m.f. proportional to the mean magnetic field ${\bf{ {\mathcal E}}}=\alpha_h \overline{{\bf{H}}}$. The coefficient $\alpha_h $ is proportional to the average helicity of the velocity field $\alpha_h \sim \overline{{\bf{v}}\textrm{rot}{\bf{v}}}$ and it is called the $\alpha $-effect in the literature \cite{3s}. The large-scale generation is usually called $\alpha_h^2$-dynamo in the absence of differential rotation, and it is called   $\alpha_h\Omega$-dynamo in the presence of rotation. The most developed at the moment is the dynamo theory in the kinematic formulation \cite{2s}-\cite{10s}, i.e. when the magnetic energy remains small compared to the kinetic energy of motion of the medium and magnetic forces have almost no effect on the flow of the medium. It is clear, that applicability of the kinematic theory of magnetic and vortex dynamos is limited. During a rather long period of time, amplified fields (vortex and magnetic ones) start influencing the flows. In this case, the behavior of the magnetic field and the motion of the substance must be considered concordantly, i.e. in the scope of the nonlinear theory. The observed magnetic fields of real objects obviously exist just in a nonlinear regime and this testifies to the significance of nonlinear theory. Simple models of nonlinear saturation of the mean field are based on the suppression of the   coefficient due to the approach of the mean field energy to equipartition with the kinetic energy of small-scale motions. In \cite{11s}, the nonlinear theory of the magnetic dynamo was constructed based on a generalization of the mean field theory (see, for example, \cite{6s}) taking into account nonlinear effects. Another alternative for constructing a nonlinear dynamo theory is the method of multiscale asymptotic expansions \cite{12s}. This method lets to strictly isolate from the entire hierarchy of perturbations the main order in which the instability arises. It should be noted that the question of the energy sources that drive the electrically conductive fluid is very important for starting the dynamo mechanism. In the papers of Busse \cite{13s}- \cite{16s}, the thermal convection of  electrically conductive fluid is considered as such a source. The possibility of simultaneously taking into account different sources of energy driving the dynamo mechanism was pointed out in \cite{17s}. 

Of no less importance are issues related to the self-excitation of magnetic fields by external e.m.f. As shown by Schl\"{u}ter and Birman \cite{18s}, the inhomogeneity of the chemical composition of a space object can lead to non-parallelism of the gradients of the electron pressure  $\nabla p_e$ and electron concentration $\nabla n_e$. The result is a "battery" electric field ${\bf E}^{(i)}=(1/en_e)\nabla p_e$, which leads to the excitation of magnetic fields $\partial {\bf B}/\partial t\approx \textrm{rot}{\bf E}^{(i)}$. Such a mechanism is associated with the generation of initial magnetic fields, which at some moment were completely absent. In this case, it is not always rightly assumed that only weak initial fields arise \cite{4s}, which are necessary to turn on the dynamo. Numerical estimates carried out in \cite{19s} showed that the magnetic fields of the Earth and planets are created by thermoelectric currents that flow in a highly conductive region inside the planet. The temperature gradients $\nabla T$  and the gradient of the thermoelectromotive force coefficient  $\nabla \alpha$ must be directed at an angle to each other. Then, the magnetic field is excited $\partial {\bf B}/\partial t\approx [\nabla T \times \nabla \alpha]$ by analogy with the "battery" effect of Birman-Schl\"{u}ter because of the non-parallelism of the vectors $\nabla T$ and $\nabla \alpha$.

Ideas of generating magnetic fields using thermoelectric currents were also pronounced in earlier works. Elsasser \cite{20s} suggested that thermoelectric power arises directly in Earth's core due to the temperature difference between the ascending and descending convective flows. In another work Runcorn \cite{21s} suggested that the thermoelectric power arises at the boundary between the mantle and the core. The theory of Earth's magnetic field created by a thermoelectric current flowing in Earth's core by the Nernst effect was developed by Hibberd \cite{22s}. In the work of Kirko \cite{23s} was proposed a physical mechanism for the occurrence of the dipole component of Earth's magnetic field by analogy with the phenomena of generation and self-excitation of a magnetic field in a fluid metal circulating in the first loop of a fast neutron reactor \cite{24s}. Here the explanation is based on the combined use of the ideas of thermoelectromagnetic hydrodynamics (TEMHD) developed by Sherkliffe \cite{25s} and the electrodynamics of mean fields \cite{6s}. A combination of thermoelectric and the  $\alpha$-effect was proposed as a mechanism for generating the magnetic field of Mercury \cite{26s}. 
In the work \cite{19s} is noted that Braginsky's conclusion \cite{17s} about the insignificance of thermopower in Earth's core is unfounded. In the paper \cite{17s} was assumed that the temperature gradient and the gradient of the thermoelectromotive force coefficient are parallel $[\nabla T_0 \times \nabla \alpha]=0$. However, as shown in this paper, under the condition  $[\nabla T_0 \times \nabla \alpha]=0$  is also possible the generation of magnetic fields due to the development of thermomagnetic instability (TMI). Thus, there are no contradictions with the Braginsky dynamo in this case. TMI was discovered in several works \cite{27s}-\cite{29s}, where the spontaneous generation of strong magnetic fields ($\sim 10^6$ G) was explained in various experiments on the interaction of laser radiation with matter in negligible times $\sim 10^{-9}$ s. A necessary condition for the development of TMI is inhomogeneity and non-isothermality of plasma. The physical mechanism of this instability is as follows. Temperature perturbations $T_1$  acting in a direction different from the initial plasma inhomogeneity lead to the excitation of a magnetic field due to the "battery" effect $\partial {\bf B}/\partial t\approx [\nabla T_1 \times \nabla n_0]$. In turn, the magnetic field affects the electronic thermal conductivity and changes the heat transfer regime. The arising heat flux supplies energy to the region with an elevated temperature promoting the growth of the initial disturbances.  

Astrophysical applications of TMI are discussed in detail in review \cite{30s}, where an explanation of the appearance of strong magnetic fields in the cores of white dwarfs, binary systems, and neutron stars is given. In a recent paper \cite{31s}, the generation of a magnetic field by TMI in the surface layers (hot plasma) of massive stars was considered. Such generation is possible in the upper layers of the atmosphere of hot stars, where deviations from local thermodynamic equilibrium form a region with an inverse temperature gradient. In \cite{31s}, the case of generation of only small-scale magnetic fields with horizontal wavelengths  $\lambda= 2\pi/k_x$ much smaller than the characteristic scale $L$ of unperturbed quantities  $\lambda \ll L$ was considered.  Thus, TMI is an alternative mechanism of the origin of magnetic fields in various astrophysical objects along with the existing theories of the turbulent \cite{2s}-\cite{10s} and convective dynamo \cite{13s}-\cite{16s}, \cite{32s}.

In some recent works \cite{33s}-\cite{35s}, the TMI in nonuniformly rotating plasma media (hot galactic disks, accretion disks) in an external axial magnetic field was considered. These papers present an analysis of the linear stability of ionized hot disks with a temperature gradient and an external axial magnetic field. Earlier, Gurevich and Helmont \cite{36s} studied the destabilizing effect of the temperature gradient on the propagation of Alfven waves in astrophysical plasma in the absence of hydrodynamic motion. As shown in \cite{33s}-\cite{35s}, the hydromagnetic and thermomagnetic effects associated with the Nernst effect can lead to amplification of waves and make disks unstable. The regimes under which both thermomagnetic and magnetorotational instabilities (MRI) can operate were discussed. MRI arises when a weak axial magnetic field destabilizes the azimuthal differential rotation of the plasma, and when the condition  $d \Omega^2/dR <0$  for the case of a nondissipative plasma is satisfied \cite{37s}. Since this condition is also satisfied for Keplerian flows $\Omega \sim R^{-3/2}$, the MRI is the most likely source of turbulence in accretion disks \cite{38s}. It was noted in \cite{35s} that even in the absence of MRI, TMI due to the Nernst effect is a good candidate for ensuring the onset of turbulence in disks. 

Unlike papers \cite{31s},\cite{33s}-\cite{35s}, in this paper, we studied the spontaneous generation of a magnetic field by TMI in a nonuniformly rotating layer of an electrically conducting fluid in the presence of an external constant axial magnetic field. Here TMI is due to the collinear temperature gradient and the gradient of the thermoelectromotive force coefficient. This work is a continuation of the research begun in \cite{39s}, where the problem of the stability of an electrically conducting fluid between two rotating cylinders (Couette flow) and the Rayleigh-Benard problem in an external constant magnetic field was considered. In contrast to \cite{39s}, we studied the influence of TM effects on convective processes, as well as the weakly nonlinear evolution of the toroidal magnetic field generated by TMI.
The work consists of the following sections. In section 1 (Introduction) gave a brief overview of the main problems of modern dynamo theory. In section 2, the basic equations of magnetohydrodynamics taking into account thermomagnetic phenomena in the Boussinesq approximation are obtained. These equations describe the generation of magnetic fields in a nonuniformly rotating electrically conducting fluid in a constant magnetic field. In section 3, the stationary magnetic convection (the Rayleigh-Benard problem) in a nonuniformly rotating liquid layer is considered, where a toroidal magnetic field is generated due to the temperature difference and specific thermopower at the layer boundaries. In section 4, we investigated  the weakly nonlinear stage of stationary inhomogeneously rotating magnetic convection taking into account TM effects. We obtained the nonlinear Ginzburg-Landau equation applying the method of perturbation theory in the small parameter of the supercriticality of the Rayleigh number $\epsilon=\sqrt{(\textrm{Ra}-\textrm{Ra}_c)/\textrm{Ra}_c}$. In the last section 5 (Conclusion) we presented the main results obtained in the work.

 \section{ Statement of the problem and equations of evolution of small perturbations}

Let us consider the following problem statement. Let us a nonuniformly rotating electrically conducting fluid (for example, iquid metal or plasma) is in a constant gravitational $\bf{g}$ and magnetic field ${\bf B}_0 $ at a constant vertical temperature gradient $\nabla T_0= \textrm{const}=-A\bf{e}$   ( $A>0$ is a constant gradient, ${\bf{e}}$ is a unit vector directed vertically upward along the axis) and a gradient of the thermoelectromotive force coefficient $\nabla\alpha \,\|\,{\bf{e}}$.  In the model considered here, we assume that the gradient of the specific thermoelectric power is associated with the inhomogeneity of the chemical composition of the conducting liquid. In this model, we assume that the gradient of the thermoelectromotive force coefficient $\nabla\alpha$  is associated with the inhomogeneity of the chemical composition of the conducting fluid. It is known that expressions for Ohm's law and heat flux  ${\bf q}$ in the presence of a magnetic field ${\bf B}$ and a temperature gradient $\nabla T $ are modified taking into account thermomagnetic phenomena \cite{40s}:
\begin{equation} \label{eq1}
{\bf E}+[{\bf V} \times {\bf B}]=\frac{{\bf j}}{\sigma}+\alpha\nabla T+{\mathcal R}[{\bf B}\times {\bf j}]+{\mathcal N}[{\bf B}\times\nabla T]
\end{equation}
\begin{equation} \label{eq2}
{\bf q}-\varphi {\bf j} = - \kappa \nabla T+\alpha T {\bf j}+{\mathcal N}T[{\bf B}\times{\bf j}]+{\mathcal L}[{\bf B}\times\nabla T], 
\end{equation}	
where $\mathcal R, \mathcal N, \mathcal L$ are the Hall, Nernst, and Leduc-Righi coefficients, respectively; $\varphi$ -- electrical potential. In expressions  (\ref{eq1})-	(\ref{eq2}), we neglected the anisotropy of the coefficients of electrical conductivity $\sigma_{\|}\approx\sigma_{\bot}=\sigma$, thermal conductivity $\kappa_{\|}\approx\kappa_{\bot}=\kappa$, and thermoelectromotive force $\alpha_{\|}\approx\alpha_{\bot}=\alpha$  due to the weakness of the external magnetic field ${\bf B}_0 $ because the parameter is small  $\beta=B_0^2/2\mu P_0 \ll 1$ ( $P_0$ is the stationary pressure of the fluid,  $\mu$  is the coefficient of magnetic permeability).  
By applying the operation  $\textrm{rot}$ to  Ohm's law (\ref{eq1}), we obtain the equation for the magnetic field induction ${\bf B} $. After substitution of expression (\ref{eq2}) into the heat balance equation $$ \rho_0 c_p\frac{dT}{d t}=- {\rm div} {\bf{q}},$$ 
let us write the equations of magnetohydrodynamics for a viscous incompressible fluid in the Boussinesq approximation taking into account thermomagnetic phenomena: 
 \[\frac{\partial {\bf{V}}}{\partial t} +({\bf{V}}\cdot\nabla ){\bf{V}}=-\frac{1}{\rho_{0} } \nabla (P+\frac{{\bf {B}}^{2} }{2\mu } )+\frac{1}{\rho_{0}\mu} {({\bf{B}}\cdot\nabla )\bf{B}}+\]
\begin{equation} \label{eq3}
+{\bf{e}}g\beta_T T +\nu \nabla^2 \bf{V} \end{equation} 
\[\frac{\partial {\bf{B}}}{\partial t} +({\bf{V}}\cdot\nabla ){\bf{B}}-({\bf{B}}\cdot\nabla ){\bf{V}}=\eta \nabla^2 {\bf{B}}-[\nabla\alpha\times\nabla T]-\]
\begin{equation} \label{eq4} 
-\frac{\mathcal R}{\mu}\textrm{rot}[{\bf{B}}\times \textrm{rot}{\bf{B}}]-\mathcal N \textrm{rot}[{\bf{B}}\times\nabla T] \end{equation}
\[\frac{\partial T}{\partial t} +({\bf{V}}\cdot\nabla ) T=-\frac{1}{\rho_0 c_p}\textrm{div}\left(- \kappa \nabla T+\frac{{\bf j}^2}{\sigma}+ \alpha T {\bf j}+\right.\frac{}{} \]
\begin{equation} \label{eq5}
\left.\frac{}{} +{\mathcal N}T[{\bf B}\times{\bf j}]+{\mathcal L}[{\bf B}\times\nabla T]\right) \end{equation} 
\begin{equation} \label{eq6} \textrm{div}~{\bf{B}}=0, \quad \textrm{div}~{\bf{V}}=0, \end{equation}
where  $\beta_T $  is the coefficient of thermal expansion, $\rho_{0} =\textrm{const}$  is the density of the medium, $\nu $ is the coefficient of kinematic viscosity,  $\eta=1/\mu\sigma$  is the coefficient of magnetic viscosity. Equation (\ref{eq4}) contains a source of excitation of a magnetic field of non-electromagnetic nature $[\nabla\alpha\times\nabla T]$, which  is an analog of the "battery" Birmann-Schl\"{u}ter effect in the plasma. The drift of the lines of force of the magnetic field in equation (\ref{eq4}) is associated not only with the movement of the fluid ${\bf{V}}$ but also with the heat flux where the rate of thermal drift is equal to: ${\bf V}_T=\mathcal N \nabla T$. The drift of the magnetic field due to the Nernst effect contributes to its penetration to a large area of the medium. Let us estimate the excited magnetic field in the stationary regime without taking into account the drift of the field and the Hall effect. Then from (\ref{eq4}) for the  $\phi$-component of the (toroidal) magnetic field, we obtain: $B_{\phi}^{max}\approx \alpha T\mu\sigma(L_B/L_\alpha)$, where $L_B$ is the characteristic scale of the excited magnetic field,   $L_\alpha$  is the characteristic scale of the medium inhomogeneity. Substituting the values of the parameters for the fluid Earth's core: $\alpha T\cong 10^{-2}$V (at temperature $T\cong 1000$K) \cite{19s}, $\mu=4\pi \cdot 10^{-7}$ V$\cdot$s/A$\cdot$m, $\sigma=3 \cdot 10^{5}$(V$\cdot$m/A)$^{-1}$ \cite{6s} with the ratio of scales $(L_B/L_\alpha)=10^{2}$, we obtain an estimated value of the toroidal magnetic field of the Earth's core $B_{\phi}^{max}\cong 10^{-1}$T, which coincides in order of magnitude with data from monograph \cite{41s}. 

Let us investigate the possibility of generating a magnetic field as a consequence of the development of TM instability by presenting all quantities in equations (\ref{eq3})-(\ref{eq6}) as the sum of the stationary and perturbed parts ${\bf V}={\bf V}_0+{\bf u}$, ${\bf B}={\bf B}_0+{\bf b}$, $P=P_0+p$, $T=T_0+\theta$. Here we assume that the stationary rotation velocity of the fluid has an azimuthal direction ${\bf V}_0 =R\Omega (R){\bf e}_\phi$. The angular velocity of rotation $\Omega (R)$ is directed vertically upward along the axis  ${\rm OZ}$. The homogeneous (constant) magnetic field  ${\bf B}_{0}\| {\bf \Omega}$ is also directed along the axis ${\rm OZ}$: ${\bf B}_0$~$=(0,0,B_0)$. Further, the magnetic field ${\bf B}_{0} $  will be called axial in the cylindrical coordinate system $(R,\phi,z)$. The stationary state of the system of equations (\ref{eq3})-(\ref{eq6}) is described by the following equations: 
 \begin{equation} \label{eq7}
\Omega ^{2} R=\frac{1}{\rho _{0} } \frac{dP_{0} }{dR},\quad \frac{1}{\rho _{0} } \frac{dP_{0} }{dz} =g\beta_T T_{0}, \quad \frac{d^2 T_0}{d z^2}=0.
\end{equation} 			
\begin{equation} \label{eq8}			
B_0\frac{d}{dz}\Omega(R)R=[\nabla\alpha\times\nabla T_0]_\phi=0			
\end{equation} 
Equations (\ref{eq7}) show that centrifugal equilibrium is established in the radial direction, and hydrostatic equilibrium in the vertical direction. From equation (\ref{eq8}) it follows that the thermoelectromotive coefficient $\alpha$ has a constant value in the radial direction: $d\alpha/dR =0$, then it can have a dependence on the coordinates $(\phi,z)$. If we consider the distribution of the chemical composition of the medium to be axisymmetric, then the condition is satisfied: $d\alpha/dz \neq 0$. In this case, the condition of collinearity of vectors is also satisfied $[\nabla\alpha\times\nabla T_0]=0$, and the gradients  $\nabla\alpha$  and  $\nabla T_0$ can be both parallel to each other $\nabla\alpha \uparrow\uparrow \nabla T_0$  and antiparallel: $\nabla\alpha \uparrow\downarrow \nabla T_0$.
The evolution equations for the perturbed quantities  $({\bf u}, {\bf b}, p, \theta)$   against the background of a stationary state take the following form: 
\[\frac{{\partial {\bf u}}}{{\partial t}} + ({\bf V}_0\cdot \nabla ){\bf u} + ({\bf u}\cdot\nabla ){\bf V}_0  =  - \frac{1}{{\rho_0 }}\nabla \left(p+\frac{1}{\mu}{\bf B}_0{\bf b}+\frac{1}{2\mu}{\bf b}^2\right) +\]
\[+ \frac{1}{{\rho _0\mu }}({\bf B}_0 \cdot\nabla ){\bf b} + g\beta_T \theta {\bf{e}}+\nu \nabla^2 {\bf u}+R_{NL}^{(1)},\]
\[ \frac{{\partial {\bf b}}}{{\partial t}} + ({\bf V}_0 \cdot\nabla ){\bf b} - ({\bf B}_0\cdot \nabla ){\bf u} - ({\bf b}\cdot\nabla ){\bf V}_0  = \eta \nabla^2 {\bf b}-[\nabla\alpha\times\nabla \theta]-\]
\begin{equation} \label{eq9}
-\frac{\mathcal R}{\mu}\textrm{rot}[{\bf{B}}_0\times \textrm{rot}{\bf{b}}]-\mathcal N \textrm{rot}([{\bf{B}}_0\times\nabla \theta]+[{\bf{b}}\times\nabla T_0])+R_{NL}^{(2)}, \end{equation}  
\[\frac{{\partial \theta}}{{\partial t}} + ({{\bf V}_0} \cdot\nabla )\theta+({\bf u}\cdot\nabla)T_0=\chi\nabla^2\theta-\frac{\alpha T_0}{\rho_0c_p\mu}({\bf K}_\alpha+{\bf K}_T)\textrm{rot}{\bf b}- \]
\[-\frac{\mathcal{N}T_0}{\rho_0c_p\mu}\textrm{div}[{\bf B}_0\times\textrm{rot}{\bf b}]-\chi_{\wedge}\textrm{div}([{\bf b}\times\nabla T_0]+[{\bf B}_0\times\nabla\theta])+R_{NL}^{(3)}, \]
\[\textrm{div}~{\bf b}=0, \quad \textrm{div}~{\bf u}=0,\]        
where the nonlinear terms $R_{NL}^{(1)}, R_{NL}^{(2)}, R_{NL}^{(3)}$ are equal respectively:
 \[ R_{NL}^{(1)}=-({\bf u}\cdot\nabla){\bf u}+\frac{1}{\rho_0\mu}({\bf b}\cdot\nabla){\bf b}, \]
\[ R_{NL}^{(2)}=({\bf b}\cdot\nabla){\bf u}-({\bf u}\cdot\nabla){\bf b}-\frac{\mathcal R}{\mu}\textrm{rot}[{\bf{b}}\times \textrm{rot}{\bf{b}}]-\mathcal N \textrm{rot}[{\bf{b}}\times\nabla \theta], \]
\[ R_{NL}^{(3)}=-({\bf u}\cdot\nabla)\theta-\frac{1}{\rho_0c_p\mu}(\alpha\nabla\theta+\theta\nabla\alpha)\textrm{rot}{\bf b}-\frac{\mathcal{N}}{\rho_0c_p\mu}(\nabla T_0\cdot[{\bf b}\times\textrm{rot}{\bf b}]+\]
\[+\nabla\theta \cdot[({\bf B}_0+{\bf b})\times\textrm{rot}{\bf b}]+(T_0+\theta)\textrm{div}[{\bf b}\times\textrm{rot}{\bf b}]+\theta\textrm{div}[{\bf B}_0\times\textrm{rot}{\bf b}])-\]
\[-\chi_{\wedge}\textrm{div}[{\bf b}\times\nabla\theta]. \]                                            
In equations (\ref{eq9}) new introduced designations  ${\bf K}_\alpha=\nabla\alpha/\alpha, {\bf K}_T=\nabla T_0/T_0$  are scales of inhomogeneity of the medium $L_\alpha\cong |{\bf K}_\alpha|^{-1}, L_T\cong |{\bf K}_T|^{-1} $,  $\chi_{\wedge}={\mathcal L}/\rho_0c_p $ is the "skewed" coefficient of thermal diffusivity.
Next, we consider the evolution of small perturbations by linearizing Eqs. (\ref{eq9}). Then we can neglect nonlinear terms. We write linearized equations (\ref{eq9}) in a cylindrical coordinate system using the following relations
$$\quad \nabla^2 \rightarrow \frac{\partial^2}{\partial z^2}+\frac{\partial^2}{\partial{R}^2}+\frac{1}{R}\frac{\partial}{\partial {R}}+\frac{1}{R^2}\frac{\partial^2}{\partial{\phi}^2}, $$
$$ (\nabla^2 {\bf u})_{R}= \nabla^2 u_R-\frac{2}{R^2}\frac{\partial }{\partial {\phi}}u_\phi-\frac{1}{R^2}u_R, $$
$$ (\nabla^2 {\bf b})_{R}= \nabla^2 b_R-\frac{2}{R^2}\frac{\partial }{\partial {\phi}}b_\phi-\frac{1}{R^2} b_R, $$
$$(\nabla^2 {\bf u})_{\phi}= \nabla^2u_\phi+\frac{2}{R^2}\frac{\partial }{\partial {\phi}} u_R-\frac{1}{R^2}u_\phi, $$     
$$ (\nabla^2 {\bf b})_{\phi}= \nabla^2 b_\phi+\frac{2}{R^2}\frac{\partial }{\partial {\phi}}b_R-\frac{1}{R^2}b_\phi. $$   
As a result, we obtain the equations of evolution of small perturbations in the linear approximation:
\[ \frac{\partial u_{R} }{\partial t} +\Omega \frac{\partial u_{R} }{\partial \phi } -2\Omega u_{\phi } -\frac{B_{0}}{\rho_{0}\mu }  \frac{\partial b_{R} }{\partial z} =-\frac{1}{\rho _{0} } \frac{\partial \widetilde p}{\partial R} +\]
\begin{equation} \label{eq10}
+\nu \left(\nabla^2 u_{R} -\frac{2}{R^2}\frac{\partial u_\phi}{\partial \phi} - \frac{u_R}{R^2} \right)\end{equation} 
\[\frac{\partial u_{\phi } }{\partial t} +\Omega \frac{\partial u_{\phi } }{\partial \phi } +2\Omega(1+\textrm{Ro})u_{R} -\frac{B_{0}}{\rho_{0}\mu } \frac{\partial b_{\phi } }{\partial z} =-\frac{1}{\rho _{0} R} \frac{\partial \widetilde p}{\partial \phi } +\]
\begin{equation} \label{eq11} 
+\nu\left(\nabla^2 u_{\phi} +\frac{2}{R^2}\frac{\partial u_R}{\partial \phi} - \frac{u_\phi}{R^2} \right)  \end{equation} 
\begin{equation} \label{eq12} \frac{\partial u_{z} }{\partial t} +\Omega \frac{\partial u_{z} }{\partial \phi } -\frac{B_{0}}{\rho_{0}\mu }\frac{\partial b_{z} }{\partial z} =-\frac{1}{\rho _{0} } \frac{\partial \widetilde p}{\partial z} + g\beta_T \theta +\nu \nabla^2 u_{z}  \end{equation} 
\[ \frac{\partial b_{R} }{\partial t} +\Omega \frac{\partial b_{R} }{\partial \phi }-B_{0}\frac{\partial u_{R} }{\partial z} =\eta \left(\nabla^2 b_{R}-\frac{2}{R^2}\frac{\partial b_\phi}{\partial\phi}-\frac{b_R}{R^2} \right)+
\alpha K_\alpha\frac{1}{R}\frac{\partial\theta}{\partial\phi}+\]
\begin{equation} \label{eq13}
+\frac{\mathcal{R}B_0}{\mu}\left(\frac{1}{R}\frac{\partial^2b_z}{\partial z\partial\phi}-\frac{\partial^2b_\phi}{\partial z^2}\right)-\mathcal{N}\left(K_T T_0\frac{\partial b_R}{\partial z}-B_0\frac{\partial^2\theta}{\partial z\partial R}\right)
\end{equation} 
\[\frac{\partial b_{\phi } }{\partial t} +\Omega \frac{\partial b_{\phi } }{\partial \phi }-2\Omega\textrm{Ro}\,b_{R}-B_{0} \frac{\partial u_{\phi } }{\partial z} =\eta\left(\nabla^2 b_{\phi} +\frac{2}{R^2}\frac{\partial b_R}{\partial \phi} - \frac{b_\phi}{R^2} \right)-\]
\begin{equation} \label{eq14} 
-\alpha K_\alpha\frac{\partial\theta}{\partial R}+\frac{\mathcal{R}B_0}{\mu}\left(\frac{\partial^2b_R}{\partial z^2}-\frac{\partial^2b_z}{\partial z\partial R}\right)-\mathcal{N}\left(K_T T_0\frac{\partial b_\phi}{\partial z}-\frac{B_0}{R}\frac{\partial^2\theta}{\partial z\partial \phi}\right)  \end{equation} 
\[\frac{\partial b_{z} }{\partial t} +\Omega\frac{\partial b_{z}}{\partial\phi }-B_{0}\frac{\partial u_{z}}{\partial z} =\eta\nabla^2 b_{z}+\]
\[+\frac{\mathcal{R}B_0}{\mu}\left(\frac{1}{R}\frac{\partial b_\phi}{\partial z}+\frac{\partial^2b_\phi}{\partial R\partial z}-\frac{1}{R}\frac{\partial^2b_R}{\partial\phi\partial z}\right)-\]
\begin{equation} \label{eq15} 
-\mathcal{N}\left(B_0\left(\frac{\partial^2\theta}{\partial R^2}+\frac{1}{R^2}\frac{\partial^2\theta}{\partial\phi^2}+\frac{1}{R}\frac{\partial\theta}{\partial R}\right)+K_T T_0\frac{\partial b_z}{\partial z} \right)  \end{equation} 
\[\frac{\partial\theta}{\partial t}+\Omega\frac{\partial\theta}{\partial \phi}-u_{z}A=\chi \nabla^2 \theta+\left(\frac{1}{R}\frac{\partial(R b_\phi)}{\partial R}-\frac{1}{R}\frac{\partial b_R}{\partial\phi}\right)\times\]
\begin{equation} \label{eq16}
\times\left(\chi_{\wedge}K_T T_0+\frac{\alpha T_0}{\rho_0c_p\mu}(K_T-K_\alpha)\right)-\frac{\mathcal{N}T_0B_0}{\rho_0c_p\mu}\nabla^2b_z  \end{equation} 
Here $ \widetilde p=p+\frac{1}{{\mu}}({\bf B}_0\cdot{\bf b})$ is the total perturbed pressure, $\chi=\kappa/\rho_0c_p$  is the coefficient of  thermoconductivity,  $c_p$  is the coefficient of specific heat, $\textrm{Ro}=\frac{R}{2\Omega} \frac{\partial\Omega}{\partial R}$  is the hydrodynamic Rossby number characterizing the inhomogeneity of rotation of the medium. Note that for solid rotation the Rossby parameter is equal to zero $\textrm{Ro}=0$, for the case of Keplerian rotation $\textrm{Ro}=-3/4$, for the Rayleigh profile of angular velocity $ \Omega(R)\sim R^{-2}$, $\textrm{Ro}=-1$ respectively. In equations (\ref{eq13})-(\ref{eq16}), we assumed that the vectors ${\bf K}_\alpha={\bf e}K_\alpha$   and   ${\bf K}_T=-{\bf e}K_T$ are antiparallel: ${\bf K}_\alpha\uparrow\downarrow{\bf K}_T$.
The system of equations (\ref{eq10})-(\ref{eq16}) is quite complicated for a complete analysis. Therefore, we consider the evolution of axisymmetric perturbations, i.e. independent of the azimuthal angle $\phi$ $(\partial/\partial\phi=0)$.
We will apply the local WKB method  for equations (\ref{eq10})-(\ref{eq16})  for perturbations  that depend on radial coordinates $R$. For this purpose, we expand all quantities in a Taylor series in the vicinity of fixed points $R_0$ leaving the terms of order zero in local coordinates  $\widetilde{R}=R-R_0$. As a result, we obtain a system of differential equations with constant coefficients taking into account the following relations
$$\Omega_0=\Omega(R_0),\quad \nabla^2 \rightarrow \widehat{D}^2+\frac{\partial^2}{\partial\widetilde{R}^2}+\frac{1}{R_0}\frac{\partial}{\partial \widetilde{R}}, \quad \widehat D \equiv \frac{\partial}{\partial z},  $$
$$ \left(\nabla^2 {\bf u} \right)_{R}= \nabla^2u_R-\frac{u_R}{R_0^2}, \;  \left(\nabla^2 {\bf b}\right)_{R}= \nabla^2 b_R-\frac{b_R}{R_0^2}, $$
$$ \left(\nabla^2 {\bf u}\right)_{\phi}= \nabla^2u_\phi-\frac{u_\phi}{R_0^2},\; \left(\nabla^2 {\bf b}\right)_{\phi}= \nabla^2 b_\phi-\frac{b_\phi}{R_0^2} . $$
All perturbations in the system of equations (\ref{eq10})-(\ref{eq16})  are represented in the form of plane waves
\begin{equation} \label{eq17}
({\bf u}, {\bf b}, \theta, \widetilde p )= \left( {\bf U}(z), {\bf H}(z), \Theta(z), \widetilde P(z)\right) \exp (\gamma t+ik\widetilde{R})
\end{equation} 
After substituting (\ref{eq17}) into the system of equations (\ref{eq10})-(\ref{eq16}),  we obtain the following equations in the short-wave approximation $ k \gg \frac{1}{R_0}$  neglecting the terms  $\frac{ik}{R_0}-\frac{1}{R_0^2} $
	\begin{equation} \label{eq18}
 \frac{d^2U_R }{dz^2} +\frac{{B_{0}}}{{\rho_0\mu\nu }}\frac{d H_R}{dz}+ \left(\frac{{2\Omega_0}}{\nu}\right)U_\phi-k_\nu^2U_R- \frac{{ik}}{{\nu\rho_0}}\widetilde P = 0 
\end{equation}
\begin{equation} \label{eq19}
 \frac{d^2U_\phi }{dz^2} +\frac{{B_{0}}}{{\rho_0\mu\nu }}\frac{d H_\phi}{dz}-\left(\frac{{2\Omega_0}}{\nu}\right)(1+\textrm{Ro})U_R-k_\nu^2U_\phi= 0 
\end{equation}
\begin{equation} \label{eq20}
 \frac{d^2U_z}{dz^2}  + \frac{{B_{0}}}{{\rho_0\mu\nu }}\frac{d H_z}{dz}-k_\nu^2U_z+\frac{g\beta_T}{\nu} \Theta-\frac{1}{{\nu\rho_0}}\frac{d\widetilde P}{dz} = 0 
\end{equation}
\[ \frac{d^2H_R }{dz^2}-\frac{\mathcal{R}B_0}{\eta\mu}\frac{d^2H_\phi }{dz^2}-\frac{\mathcal{N}K_TT_0}{\eta}\frac{dH_R}{dz}+\frac{\mathcal{N}B_0ik}{\eta}\frac{d\Theta}{dz}+\]
\begin{equation} \label{eq21}
+\frac{{B_{0}}}{{\eta}}\frac{d U_R}{dz}-k_\eta^2H_R= 0 
\end{equation}
\[\frac{d^2H_\phi}{dz^2}-\frac{\mathcal{R}B_0}{\eta\mu}\frac{d^2H_R}{dz^2}+\frac{ik\mathcal{R}B_0}{\eta\mu}\frac{d H_z}{dz}-\frac{\mathcal{N}K_TT_0}{\eta}\frac{dH_\phi}{dz}+\]
\begin{equation} \label{eq22}
 +\frac{{B_{0}}}{{\eta}}\frac{d U_\phi}{dz}+\left(\frac{{2\Omega_0}}{\eta}\right)\textrm{Ro}H_R-k_\eta^2H_\phi+\frac{ik\alpha}{\eta }K_\alpha\Theta= 0 \end{equation}
\[\frac{d^2H_z}{dz^2}+\frac{\mathcal{R}B_0ik}{\eta\mu}\frac{d H_\phi}{dz}+\frac{\mathcal{N}B_0}{\eta}k^2\Theta
+\frac{\mathcal{N}K_TT_0}{\eta}i\left(k H_R\right)+\]
\begin{equation} \label{eq23}
+\frac{{B_{0}}}{{\eta}}\frac{d U_z}{dz}-k_\eta^2H_z=0
\end{equation}
\[\frac{d^2\Theta}{dz^2}+\frac{\mathcal{N}T_0B_0ik}{\rho_0c_p\mu\chi}\frac{d H_R}{dz}+ik H_\phi\frac{s}{\chi}+\]
\begin{equation} \label{eq24}
+\frac{\mathcal{N}T_0B_0}{\rho_0c_p\mu\chi} k^2H_z-\frac{K_TT_0}{\chi}U_z-k_\chi^2\Theta=0
\end{equation}
Here the following notation is introduced 
\[ k_{\nu}^2= \frac{{\gamma}}{\nu} + k^2,\; k_{\eta}^2=\frac{{\gamma}}{\eta} + k^2,\; k_{\chi}^2=\frac{{\gamma}}{\chi} + k^2, \]
\[  s=\frac{\alpha T_0}{\rho_0c_p\mu}\left[\left(\frac{\mu \mathcal{L}}{\alpha}+1\right)K_T-K_\alpha\right].   \]                           
Equations (\ref{eq18})-(\ref{eq24}) were supplemented by equations of the solenoidality of the fields  ${\bf u}$ and  ${\bf b}$:
\begin{equation} \label{eq25}
\frac{dU_z}{dz}+ikU_R=0,   \quad  \frac{dH_z}{dz}+ikH_R=0. \end{equation}
Let us continue with a more detailed analysis of Eqs. (\ref{eq18})-(\ref{eq24}).

\section{Generation of a magnetic field by thermomagnetic effects in a thin layer of a nonuniformly rotating fluid}

\begin{figure}
  \centering
	\includegraphics[width=10 cm, height=6 cm]{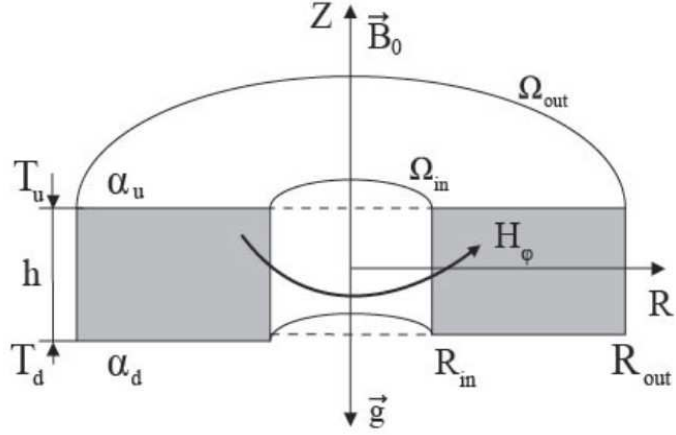} \\
\caption{An electrically conductive fluid fills in a layer between two rotating cylinders with angular velocities $\Omega_{{in}}$  and $\Omega_{{out}}$ , respectively. The lower surface of the layer has the temperature  $T_d$  and the thermoelectromotive force coefficient $\alpha_d$, and the upper surface has $T_u$  and $\alpha_u$: $T_{d}>T_{u}$, $\alpha_d<\alpha_u$. The generated magnetic field has an azimuthal direction $H_\phi$. }\label{fg1}
\end{figure}

Let us consider a stationary flow of a nonuniformly rotating incompressible viscous electrically conductive fluid, which is modeled by the Couette-Taylor flow enclosed between two rotating cylinders with an angular velocity of rotation $\Omega(R)$:
\[ \Omega(R)=\frac{\Omega_2R_2^2-\Omega_1R_1^2}{R_2^2-R_1^2}+\frac{(\Omega_1-\Omega_2)R_1^2R_2^2}{R^2(R_2^2-R_1^2)}, \]                            
where $R_1=R_{{in}}, R_2=R_{{out}}, \Omega_1=\Omega_{{in}}, \Omega_2=\Omega_{{out}}$  are radius and angular velocity of rotation of the inner and outer cylinder, respectively. The choice of this type of flow is due to the possibility of realization of the theory developed here in laboratory experiments. The height of the cylinders corresponds to a liquid layer of finite thickness  $h$  under the condition  $h \ll (R_{{out}}-R_{{in}})$.
The geometry of the problem is shown in Fig. \ref{fg1}. On the lower plane of the layer is supported  a higher temperature  $T_{d}$  than on the upper plane $T_{u}$: $T_{d}>T_{u}$ -- the heating from below. The  thermoelectromotive force coefficient  $\alpha_d$  on the lower ("hot") plane is less than on the upper ("cold") plane $\alpha_u$: $\alpha_d<\alpha_u$. This situation is quite possible if we take into account the dependence of the thermoelectromotive force coefficient  on temperature  $\alpha \sim \psi/T_0$ ( $\psi$  is the chemical potential)  \cite{40s}. A spatially inhomogeneous distribution inside a layer $T_0(z)$ and $\alpha(z)$ can be represented as a linear dependence on $z$: $$T_0(z)=T_d-\frac{\Delta T}{h}\cdot z,\; \Delta T= T_d-T_u, \quad \alpha(z)= \alpha_d+\frac{\Delta\alpha}{h}\cdot z, \; \Delta\alpha= \alpha_u-\alpha_d . $$ 
As is known \cite{42s},  the temperature difference at the boundaries of the layer in the gravity field ${\bf g}$  leads to a violation of mechanical equilibrium in the system.  In this case, the convective instability develops and convective cells are formed. Temperature perturbations acting in the radial direction lead to the appearance of a vortex thermal current due to the difference in the values of the thermoelectromotive force coefficient $\Delta\alpha(z)$ at the layer boundaries. This current induces disturbances of the azimuthal (toroidal) magnetic field $H_\phi$  (see Fig. \ref{fg1}), which influences the heat transfer regime. The excited magnetic field  $H_\phi$ creates heat fluxes directed perpendicular to the field itself and the temperature gradient. Thus,  positive feedback is established: newly arising heat fluxes create a vortex thermoelectromotive power, which enhances magnetic field disturbances $H_\phi$. Naturally,  the thermomagnetic instability will affect to the development of convective instability. We will consider this issue in current section.

\subsection{ Dispersion equation for TM perturbations}

We use the system of equations (\ref{eq18})-(\ref{eq25}) to describe magnetic convection in a thin layer of a nonuniformly rotating fluid taking into account thermogalvanomagnetic effects. For analysis of the system of equations (\ref{eq18})-(\ref{eq25}), it is convenient to bring it to a dimensionless form by introducing dimensionless quantities, which we have noted with an asterisk:
\[\left( {R_0}^ * , z^ * \right) = h^{-1} (R_0,z), \left( {U_R ^ *  ,U_\phi ^ *  ,U_z ^ *  } \right) = \left(\frac{\chi}{h}\right)^{-1} (U_R ,U_\phi  ,U_z ),\]
\[ \left( {H_R ^ *  ,H_\phi ^ *  ,H_z ^ *  } \right) = B_0^{ - 1} (H_R ,H_\phi  ,H_z ), \]
\[ \Theta^ *   = \Theta (\Delta T)^{-1},\quad \widetilde{ P}^ *   = \widetilde{P}\left( {\frac{{h^2}}{{\rho_0 \nu \chi }}} \right) ,\quad t^* = t\left( {\frac{\nu}{{h^2}}} \right), \quad \gamma^*  = \frac{{h^2}}{\nu}\gamma.\]
Omitting the asterisk icon, we have got the following system of dimensionless equations
\begin{equation} \label{eq1m}
\widehat L_\nu  U_R +\Pr\textrm{Pm}^{-1}\textrm{Ha}^2 \widehat DH_R+\sqrt{\textrm{Ta}} U_\phi-i k \widetilde P = 0 \end{equation}
\begin{equation} \label{eq2m}
\widehat L_\nu  U_\phi +\Pr\textrm{Pm}^{-1}\textrm{Ha}^2\widehat DH_\phi-\sqrt{\textrm{Ta}}(1+\textrm{Ro})U_R  = 0 \end{equation}
\begin{equation} \label{eq3m}
\widehat L_\nu  U_z + \Pr\textrm{Pm}^{-1}\textrm{Ha}^2 \widehat DH_z  +\textrm{Ra}\Theta-\widehat D\widetilde P = 0 \end{equation}
\begin{equation} \label{eq4m}
\widehat L_\eta H_R+{\Pr}^{-1}\textrm{Pm}\widehat DU_R+\textrm{R}_{\mathcal{N}}(\widehat D H_R+ik \widehat D \Theta)-\textrm{R}_{{H}}\widehat D^2 H_\phi= 0 \end{equation}
\[\widehat L_\eta H_\phi + {\Pr}^{-1}\textrm{Pm}\widehat DU_\phi+\sqrt{\textrm{Ta}}\textrm{PmRo}H_R+\]
\begin{equation} \label{eq5m}
+\textrm{R}_{{H}}(ik\widehat D H_z-\widehat D^2 H_R)+\textrm{R}_{\mathcal{N}}\widehat D H_\phi+ik\textrm{R}_{\alpha}\Theta = 0 \end{equation}
\begin{equation} \label{eq6m}
\widehat L_\eta H_z + {\Pr}^{-1}\textrm{Pm}\widehat DU_z+\textrm{R}_{{{H}}}ik\widehat D H_\phi+\textrm{R}_{\mathcal{N}}(k^2\Theta-ikH_R)= 0 \end{equation}
\begin{equation} \label{eq7m}
\widehat L_\chi \Theta+U_z+q_{\mathcal{N}}(ik\widehat D H_R+k^2H_z)-q_\alpha ikH_\phi= 0  \end{equation}
\begin{equation} \label{eq8m}
\widehat D U_z+ik U_R=0,\;\widehat D H_z+ik H_R=0     \end{equation}
\[\quad \widehat L_\nu=\widehat D^2-\gamma-k^2, \quad \widehat L_\eta= \widehat D^2-\textrm{Pm}\gamma-k^2, \]
\[ \widehat L_\chi=\widehat D^2-\Pr \gamma- k^2,\quad \widehat D\equiv\frac{d}{dz}, \]
where $\textrm{Pr}=\nu/\chi$ is the Prandtl number, $\textrm{Pm}=\nu/\eta$ is the Prandtl magnetic number, $\textrm{Ta}=\frac{4{\Omega_0}^2 h^4}{\nu^2}$ is the Taylor number, $\textrm{Ha}=\frac{B_0 h}{\sqrt{\rho_0\mu\nu\eta}}$ is the Hartman number, $\textrm{Ra}=\frac{g\beta_T (\Delta T) h^3}{\nu\chi}$ is the Rayleigh number, dimensionless parameters:  $ \textrm{R}_{{H}}= \frac{\mathcal{R}B_0}{\mu\eta}$ is the Hall number, $\textrm{R}_{\mathcal{N}}=\frac{\mathcal{N}\Delta T}{\eta}$ is the Nernst number, $\textrm{R}_{\alpha}=\frac{\Delta\alpha\Delta T}{\eta B_0}$  is  the thermoelectromotive force number.  Dimensionless parameters
 $$q_{\mathcal{N}}=\frac{\mathcal{N}T_0B_0^2}{\rho_0c_p\mu\chi(\Delta T)} \quad \textrm{and} \quad  q_\alpha=\frac{\alpha T_0B_0}{\rho_0c_p\mu\chi(\Delta T)}\left[\left(\frac{\mu\mathcal{L}}{\alpha}+1\right)\frac{\Delta T}{T_0}+\frac{\Delta\alpha}{\alpha}\right] $$     
are associated with the influence of the Nernst, thermoelectromotive force, and Leduc-Righi effects on the heat transfer process. We use the Chandrasekhar numbers  $\textrm{Q}=\textrm{Ha}^2$  and  $\widetilde{\textrm{Q}}=\textrm{Q}\textrm{Pm}^{-1}\textrm{Pr} $ 
instead of the Hartmann number  $\textrm{Ha}$  for convenience. The generation of magnetic fields arises due to the effects associated with the inhomogeneity of the thermoelectromotive force coefficient and "magnetization" of the heat flux (Leduc-Righi effect). Then, in equation (\ref{eq9m}), we restrict ourselves to only taking these effects into account. As a result of simple but cumbersome mathematical operations, equations (\ref{eq1m})-(\ref{eq8m}) are reduced to one differential equation for $U_z$: 
\[[ \widehat a_{11} \left( {\widehat a_{22}\widehat a_{33}-\widehat a_{23} \widehat a_{32} }\right) + \widehat a_{12} \left(\widehat a_{23} \widehat a_{31}-\widehat a_{21} \widehat a_{33} \right)+\]
\begin{equation} \label{eq9m}
\widehat a_{13} \left(\widehat a_{21} \widehat a_{32}-\widehat a_{31} \widehat a_{22} \right) ]U_z  = 0, \end{equation} 
where the explicit form of the operators $\widehat a_{ik}$ $(i=1,2,3;  k=1,2,3)$  is given in the Appendix $\textrm{A}$.
Equation (\ref{eq9m}) is supplemented with boundary conditions only in the  $z$-direction
 \begin{equation} \label{eq10m}
 U_z=\frac{d^2 U_z}{dz^2}=0, \quad \textrm{at} \quad z=0, \quad \textrm{and} \quad z=1. \end{equation}          
Equation (\ref{eq9m}) with boundary condition (\ref{eq10m}) describes convective phenomena in a thin layer of a nonuniformly rotating magnetized fluid with thermomagnetic effects. For simplicity, the solution of equation (\ref{eq9m}) with boundary condition (\ref{eq10m}) will be sought in the form of a single-mode approximation 
\begin{equation} \label{eq11m} U_z=W_0 \sin \pi z,
\end{equation}                                                  	
where $W_0$ is a constant amplitude. Substituting (\ref{eq11m}) into (\ref{eq9m}) and integrating over the layer thickness $z=(0,1)$, we obtain the dispersion equation
 \begin{equation} \label{eq12m}
\textrm{Ra} = \left(\textrm{Ra}^{(0)}-\textrm{Ra}^{(TM)}\right)\left(1-\frac{k^2q_\alpha R_\alpha}{\Gamma_\eta\Gamma_\chi}\right)\Delta^{-1},\end{equation}                                 	
where $\textrm{Ra}^{(0)}$ is the contribution to the dispersion equation without taking into account TM effects, obtained in \cite{39s}:
 \[\textrm{Ra}^{(0)} = \frac{{ \Gamma_\chi(a^2\Gamma_A^4  +\pi^2\textrm{Ta}(1+\textrm{Ro})\Gamma_\eta^2  + \pi^4\textrm{Ha}^2\textrm{TaRoPm})}}{{k^2\Gamma_\eta\Gamma_A^2 }}, \] 
\[ \textrm{Ra}^{(TM)}=q_\alpha R_\alpha\frac{a^2(\gamma+a^2)}{\Gamma_\eta}+q_\alpha R_\alpha\frac{\pi^2\textrm{Ta}}{\Gamma_A^2}(1+\textrm{Ro})+R_\alpha\frac{\pi^2\textrm{Q}}{\Gamma_A^2}\textrm{Pm}\textrm{Pr}^{-1}\sqrt{\textrm{Ta}},\]
\[\Delta=\left(1-\frac{q_\alpha}{\Gamma_\eta^2}\pi^2\textrm{Pm}^2\textrm{Pr}^{-1}\textrm{Ro}\sqrt{\textrm{Ta}}\right)\left(1-\frac{q_\alpha R_\alpha}{\Gamma_\chi\Gamma_A^2}k^2(\gamma+a^2)\right)+ \frac{q_\alpha\pi^2\textrm{Pm}\textrm{Pr}^{-1}}{\Gamma_\eta^2\Gamma_\chi\Gamma_A^2}\times   \]\[\times\left(\sqrt{\textrm{Ta}}(1+\textrm{Ro})\Gamma_\eta(\Gamma_\eta\Gamma_\chi-k^2q_\alpha R_\alpha)+\pi^2\textrm{QPmRo}\sqrt{\textrm{Ta}}\Gamma_\chi\right)-\frac{q_\alpha R_\alpha \pi^2k^2\textrm{Q}}{\Gamma_\eta\Gamma_\chi\Gamma_A^2},\]             
where the new notation is introduced
\[\Gamma_A^2=(\gamma+a^2)(\gamma\textrm{Pm}+a^2)+\pi ^2\textrm{Ha}^2,\quad \Gamma_\chi=\gamma\Pr+a^2, \] 
\[\Gamma_\eta=\gamma\textrm{Pm}+a^2,\quad a^2=\pi^2+k^2.\]
In the absence of thermal processes, MRI arises in a nonuniformly rotating layer of an electrically conductive fluid in a constant magnetic field. In this case, equation (37) coincides with the dispersion equation for the standard MRI (SMRI) taking into account dissipative processes \cite{43s}. The threshold value of the hydrodynamic Rossby number Ro is determined using the condition $\gamma=0$ and has the form:
\[\textrm{Ro}_{{cr}}  =-\frac{a^2(a^4+\pi^2 \textrm{Ha}^2)^2+\pi^2a^4 \textrm{Ta}}{\pi^2\textrm{Ta}(a^4+\pi^2 \textrm{Ha}^2 \textrm{Pm})}.\] When transited to dimensional variables
 \[\frac{\pi^2 \textrm{Ha}^2}{a^4} \rightarrow \frac{\omega_A ^2}{\omega_\nu \omega_\eta},\; \frac{\pi^2 \textrm{Ha}^2\textrm{Pm}}{a^4} \rightarrow \frac{\omega_A ^2}{ \omega_\eta ^2},\; \frac{\textrm{Ta}}{a^4} \rightarrow \frac{4\Omega^2}{\omega_\nu^2},\; \frac{\pi^2}{a^2}\rightarrow \xi^2\]
 the expression for ${\rm Ro}_{cr}$ is found \cite{43s}
\[\textrm{Ro}_{{cr}}=-\frac{{(\omega _A^2+\omega_\nu\omega_\eta)^2+4\xi^2\Omega^2 \omega_\eta^2 }}{{4\Omega^2 \xi^2 (\omega_A^2 +\omega_\eta^2 )}},  \]
The criterion for MRI appearance is the condition imposed on the angular velocity profile $\Omega(R)$ of the rotating liquid, i.e., Rossby number $\textrm{Ro}>\textrm{Ro}_{{cr}}$. Let us now analyze a more general case when there is the heating of the fluid layer $\textrm{Ra}\neq 0$  and its nonuniform rotation $\textrm{Ro}\neq 0 $ taking into account the thermomagnetic effects.

\subsection{Stationary convection regime}

Obviously, for the stationary convection mode, the increment $\gamma$ is zero $(\gamma=0)$, therefore from formula (\ref{eq12m}) we can find the critical value of the Rayleigh number  $\textrm{Ra}_{st}$ for stationary convection:
\[\textrm{Ra}_{st}=\left[\textrm{Ra}_{st}^{(0)}-a^2q_\alpha R_\alpha-\frac{\pi^2R_\alpha}{a^4+\pi^2\textrm{Q}}\left(\textrm{Ta}(1+\textrm{Ro})q_\alpha+\widetilde{\textrm{Q}}\sqrt{\textrm{Ta}}\right)\right]\times\]
 \begin{equation} \label{eq13m}
\times\left[1+\frac{q_\alpha\pi^2\textrm{Pm}\textrm{Pr}^{-1}\sqrt{\textrm{Ta}}(1+\textrm{Ro}(1-\textrm{Pm}))}{a^4+\pi^2\textrm{Q}}\right]^{-1},
\end{equation}
where \[ \textrm{Ra}_{st}^{(0)}=\frac{a^6}{k^2}+\frac{a^2\pi^2\textrm{Q}}{k^2}+\frac{\pi^2\textrm{Ta}}{k^2}\cdot \frac{a^4+ \textrm{Ro}(a^4+\pi^2\textrm{Q}\textrm{Pm})}{a^4+\pi^2\textrm{Q}} \,. \]
The minimum value of the critical Rayleigh number is found from the condition  $\partial\textrm{Ra}_{st}/\partial k =0$  and corresponds to the wavenumbers $k=k_c$ that satisfy the following equation:
\[\left(1+\frac{q_\alpha\pi^2\textrm{Pm}\textrm{Pr}^{-1}\sqrt{\textrm{Ta}}(1+\textrm{Ro}(1-\textrm{Pm}))}{(\pi^2+k_c^2)^2+\pi^2\textrm{Q}}\right)\cdot\left(\textrm{M}(k_c)-\textrm{R}_{TM}\cdot\frac{k_c^3(\pi^2+2k_c^2)}{(\pi^2+k_c^2)}+\right.\]
\[\left.+\textrm{R}_{TM}\cdot\frac{2\pi^2k_c^3(\textrm{Ta}(1+\textrm{Ro})\textrm{R}_{TM}+R_\alpha\widetilde{\textrm{Q}}\sqrt{\textrm{Ta}}}{(\pi^2+k_c^2)((\pi^2+k_c^2)^2+\pi^2\textrm{Q})^2} \right)+\]
\[+\left(\frac{(\pi^2+k_c^2)^3}{k_c^2}-\textrm{R}_{TM}\cdot k_c^2(\pi^2+k_c^2)+\frac{\pi^2+k_c^2}{k_c^2}\cdot\pi^2\textrm{Q}+\right.\]
\[\left.+\frac{\pi^2\textrm{Ta}(1+\textrm{Ro})((\pi^2+k_c^2)^2-k_c^2\textrm{R}_{TM})+\pi^4\textrm{QPmRoTa}-k_c^2\pi^2R_\alpha\widetilde{\textrm{Q}}\sqrt{\textrm{Ta}}}{k_c^2((\pi^2+k_c^2)^2+\pi^2\textrm{Q})^2}\right)\times\]
\[\times\frac{2k_c^3\pi^2q_\alpha\textrm{Pm}\textrm{Pr}^{-1}\sqrt{\textrm{Ta}}(1+\textrm{Ro}(1-\textrm{Pm}))}{(\pi^2+k_c^2)((\pi^2+k_c^2)^2+\pi^2\textrm{Q})^2}=0,\] 
where $\textrm{R}_{TM}=q_\alpha R_\alpha$  is a dimensionless parameter depending on the temperature gradient and gradient of the thermoelectromotive force coefficient. In the limiting case, when TM effects are absent, this equation coincides with the result of \cite{39s}:
\[\textrm{M}(k_c)=\frac{{2k_c^2-\pi^2}}{{k_c}}-\frac{{\pi^4\textrm{Q}}}{{k_c (\pi^2+k_c^2)^2}}+\frac{{2\pi^2 k_c\textrm{Ta}(1+\textrm{Ro})}}{(\pi^2+k_c^2)\left((\pi^2+k_c^2)^2+\pi^2\textrm{Q}\right)}-\]
\[-\frac{\pi^2\textrm{Ta}((\pi^2+k_c^2)^2+\pi^2\textrm{Q}+2k_c^2(\pi^2+k_c^2))}{k_c((\pi^2+k_c^2)^2+\pi^2\textrm{Q})^2}-\]
\[-\frac{\pi^2\textrm{TaRo}((\pi ^2+k_c^2)^2+\pi^2\textrm{QPm})((\pi^2+k_c^2)^2+\pi^2\textrm{Q}+2k_c^2(\pi^2+k_c^2))}{k_c(\pi^2+k_c^2)^2((\pi^2+k_c^2)^2+\pi^2\textrm{Q})^2}= 0 \]
Let us consider some limiting cases.
\begin{enumerate}
\item In the absence of rotation  $(\textrm{Ta}=0, \textrm {Ro}=0 )$ and magnetic field $({\bf{B}}_0=0)$ from expression (\ref{eq13m}) we found
 \begin{equation} \label{eq14m}
\textrm{Ra}_{st}=\frac{(k^2+\pi^2)^3}{k^2}-\textrm{R}_{TM}(k^2+\pi^2)
\end{equation}                                      
If the gradient of the thermoelectromotive force coefficient is zero (the medium is chemically homogeneous), then the well-known result follows from (\ref{eq14m}) $\textrm{Ra}_{st}^{\textrm{min}}=(k^2+\pi^2)^3/k^2$. Here, the minimum value of the critical Rayleigh number $\textrm{Ra}_{st}=27\pi^4/4$  reaches for wavenumber  $k_c=\pi/\sqrt{2}$ \cite{42s}. The minimum value of the critical
\begin{figure}
  \centering
	\includegraphics[width=7 cm, height=7 cm]{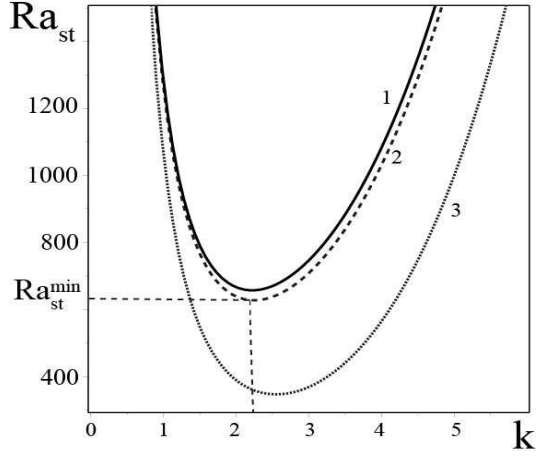}\\
	\caption{Dependence of the stationary Rayleigh number $\textrm{Ra}_{st}$ on wavenumbers  $k$ in the absence of rotation $\textrm{Ta}=0$ and magnetic field ${\bf B}_0=0$. Curve 1 corresponds to the parameter value $\textrm{R}_{TM}=0$, curve 2 -- $\textrm{R}_{TM}=2$, curve 3 - $\textrm{R}_{TM}=20$. }\label{fg2}
\end{figure}
Rayleigh number $\textrm{Ra}_{st}^{\textrm{min}}$  is calculated by the formula (\ref{eq14m})  at $k=k_c$, which satisfies the following relation
   \[ 2\left(\frac{k_c}{\pi}\right)^6+3\left(\frac{k_c}{\pi}\right)^4=1+\left(\frac{k_c^4}{\pi^6}\right)\textrm{R}_{TM} \]                        
The numerical value $\textrm{Ra}_{st}^{{min}}$ in Fig. \ref{fg2} corresponds to a point on the neutral curve separating the regions of stable and unstable disturbances. It can be seen here that with an increase in the coefficient the minimum value of the critical Rayleigh number decreases, i.e. the threshold for the development of instability decreases. A numerical estimate of the coefficient  $\textrm{R}_{TM}$ was carried out for the physical parameters of the Earth's core: $\rho_0 \approx 7\cdot 10^3$kg/m$^{3}$ is the density of molten iron, $c_p\approx 835$ J/kg$\cdot$K  is specific heat \cite{41s} and $\kappa=39$ W/m$\cdot$K is the thermal conductivity coefficient for iron in the molten state \cite{41s}. These parameters give the value of the thermal diffusivity $\chi=\kappa/\rho_0 c_p \approx 6.7\cdot 10^{-6}$ m$^{2}$/s , which turns out to be much less than the magnetic viscosity coefficient $\eta=1/\mu\sigma\approx 2.65$ m$^{2}$/s: $\eta \gg \chi$. The value $\textrm{R}_{TM}\approx 2$  was obtained for variations of the thermoelectromotive force coefficient $\Delta\alpha=3\cdot 10^{-4}$V/K  and temperature  $\Delta T=2000$K, and with the increase $\textrm{R}_{TM}\approx 20$ of the variations of the thermoelectromotive force coefficient  to $\Delta\alpha\approx 10^{-3}$V/K.
 \item If the medium rotates nonuniformly  $(\textrm{Ro}\neq 0)$ but without an external magnetic field  $({\bf{B}}_0=0)$, then expression (\ref{eq13m}) takes the form
 \[\textrm{Ra}_{st}=\frac{(k^2+\pi^2)^3}{k^2}+\frac{\pi^2\textrm{Ta}}{k^2}(1+\textrm{Ro})-\]
\begin{equation} \label{eq15m}
-\textrm{R}_{TM}\left(k^2+\pi^2+\frac{\pi^2\textrm{Ta}}{(k^2+\pi^2)^2}(1+\textrm{Ro})\right)
\end{equation}            
We also obtained the well-known result \cite{42s} for the case of a non-conductive $(\sigma=0)$  and uniformly rotating $(\textrm{Ro}=0)$  medium from expression (\ref{eq15m}):
\[\textrm{Ra}_{st}=\frac{(k^2+\pi^2)^3}{k^2} +\frac{\pi^2\textrm{Ta}}{k^2}.\]
\begin{figure}
  \centering
	\includegraphics[width=7 cm, height= 7 cm]{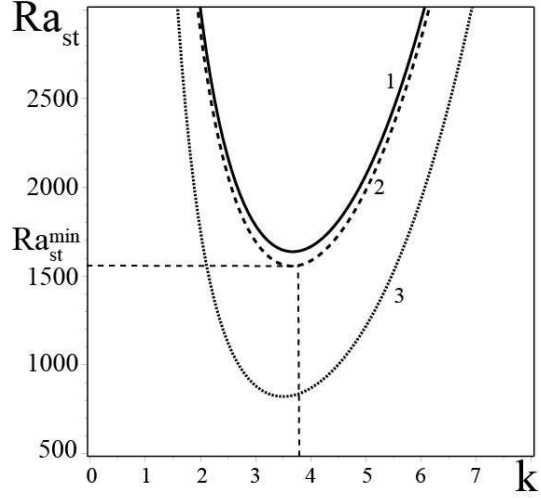}\\
	\caption{Dependence of the stationary Rayleigh number $\textrm{Ra}_{st}$ on the wavenumbers  $k$ for a medium uniformly rotating with the Taylor number  $\textrm{Ta}=946$ $(\textrm{Ro}=0)$  without an external magnetic field ${\bf B}_0=0$ . Curve 1 corresponds to the parameter value $\textrm{R}_{TM}=0$, curve 2 -- $ \textrm{R}_{TM}=2$, curve 3 --  $\textrm{R}_{TM}=20$.}\label{fg3}
\end{figure}
Similarly, we calculated the minimum value of the critical Rayleigh number $\textrm{Ra}_{st}^{{min}}$  using formula (\ref{eq15m}) at $k=k_c$ , which satisfies the following relation
\[2\left(\frac{k_c}{\pi}\right)^6+3\left(\frac{k_c}{\pi}\right)^4=1+\left(\frac{k_c^4}{\pi^6}\right)\textrm{R}_{TM}+\]
\[+\frac{\textrm{Ta}}{\pi^4}(1+\textrm{Ro})\left(1-\textrm{R}_{TM}\frac{k_c^4}{(k^2+\pi^2)^3}\right)   \]                       
In Fig. \ref{fg3} shows the dependence of the critical (stationary) Rayleigh number $\textrm{Ra}_{st}$ (\ref{eq15m}) on wavenumbers $k$ in the presence $\textrm{R}_{TM}\neq 0$ and absence $\textrm{R}_{TM}=0$ of the influence of TM effects. Here we observe that with the increase of the coefficient $\textrm{R}_{TM}$  the minimum value of the critical Rayleigh number  $\textrm{Ra}_{st}^{{min}}$, for a uniformly rotating medium with the Taylor number $\textrm{Ta}=946$, decreases, i.e. the threshold for the development of instability decreases. The Taylor number $\textrm{Ta}=946$  was calculated for the parameters of the Earth's core: $\Omega_0=4\cdot 10^{-5}$s$^{-1}$ is  the angular velocity of rotation;  $\nu=2.6$ m$^2$/s is the coefficient of hydrodynamic viscosity is considered equal to the coefficient of magnetic viscosity $\textrm{Pm}=1$; $h=10^{3}$m is the thickness of the convective layer. The estimates of the physical values of the Earth's core given in \cite{41s} have a fairly wide interval so we chose the above values of density $\rho_0$, electrical conductivity  $\sigma$, thermal diffusivity  $\chi$, viscosity ($\nu, \eta$), temperature  $T_0$, 
\begin{figure}
  \centering
	\includegraphics[width=7 cm, height= 7 cm]{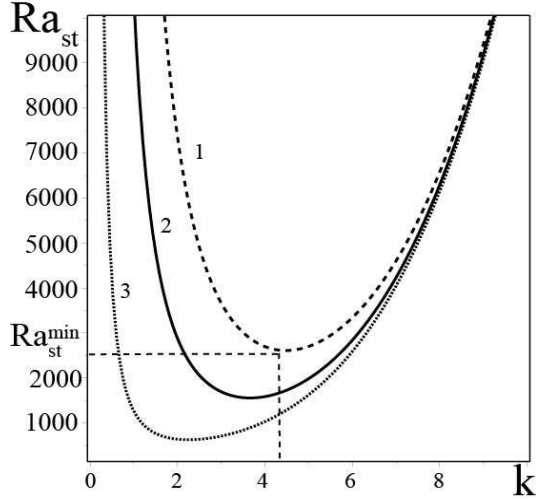}\\
	\caption{Dependence of the stationary Rayleigh number $\textrm{Ra}_{st}$  on wavenumbers $k$ for an inhomogeneously rotating medium with the Taylor number $\textrm{Ta}=946$ without an external magnetic field  ${\bf B}_0=0$ at the constant parameter $\textrm{R}_{TM}=2$. Curves 1,2,3 correspond to  Rossby numbers $\textrm{Ro}=2$,  $\textrm{Ro}=0$,  $\textrm{Ro}=-1$, respectively.}\label{fg4}
\end{figure}
thermoelectromotive force coefficient  $\alpha$, etc. convenient for numerical calculations and reasonable physical interpretation of the results. 

Next, we fix the value of the coefficient $\textrm{R}_{TM}=2$, and the Rossby number $\textrm{Ro}$  will be varied.  Fig. \ref{fg4} shows that with the increase of the positive profile of the Rossby number  $\textrm{Ro}$  the minimum value of the critical Rayleigh number also increases  $\textrm{Ra}_{st}^{{min}}$, i.e., the threshold for the development of instability increases. On the other hand, we observe the decrease of the critical Rayleigh number for negative rotation profiles $(\textrm{Ro}=-1)$ (curve 3), i.e. the threshold for the development of instability is  lower in compare to the case of uniform  $(\textrm{Ro}=0)$ (curve 2) and nonuniform  $(\textrm{Ro}=2)$ (curve 1) rotation.

\item Let us consider the case when there is no rotation $(\textrm{Ta}=0, \textrm {Ro}=0 )$  but there is an external magnetic field  $({\bf{B}}_0\neq 0)$.  Then from expression (\ref{eq13m}) we found the critical value of the Rayleigh number:
\begin{equation} \label{eq16m}
\textrm{Ra}_{st}=\frac{(k^2+\pi^2)^3}{k^2}\cdot\left(1+\frac{\pi^2\textrm{Q}}{(k^2+\pi^2)^2}\right)-\textrm{R}_{TM}(k^2+\pi^2)
\end{equation}
 If   $\textrm{R}_{TM}=0$, then we may obtain the result known from the monograph \cite{42s}. The minimum value of the critical Rayleigh number  $\textrm{Ra}_{st}^{\textrm{min}}$  is determined from formula (\ref{eq16m}) at   $k=k_c$, which satisfies the following relation 
\[ 2\left(\frac{k_c}{\pi}\right)^6+3\left(\frac{k_c}{\pi}\right)^4= 1+\frac{\textrm{Q}}{\pi^2}+\textrm{R}_{TM}\left(\frac{k_c^4}{\pi^6}\right) \]
 The graph in Fig. \ref{fg5} shows the dependence of the critical (stationary) Rayleigh number  $\textrm{Ra}_{st}$ (\ref{eq16m}) on the wavenumbers  $k$. The magnitude of the external poloidal (or meridional) magnetic field emerging from the core to the Earth's surface is of the order of $B_0=10^{-1}$T \cite{41s}, which will correspond to the Chandrasekhar
\begin{figure}
  \centering
	\includegraphics[width=7 cm, height=7 cm]{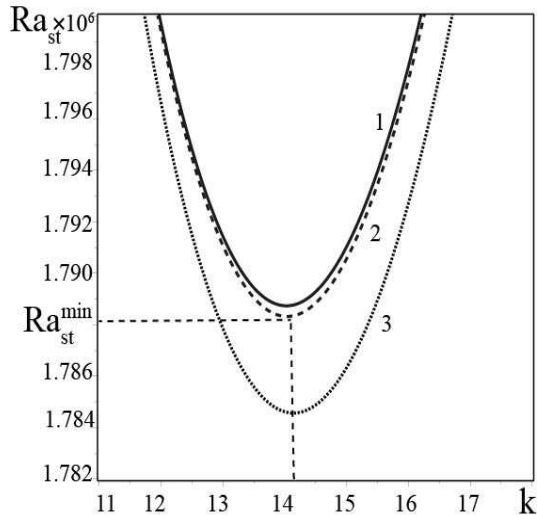}\\
	\caption{Dependence of the stationary Rayleigh number  $\textrm{Ra}_{st}$ on wavenumbers  $k$ for a non-rotating medium ($\textrm{Ta}=0$) in an external magnetic field  $B_0=10^{-1}$T at different values of the parameter  $\textrm{R}_{TM}$. Curves 1,2,3 correspond to   $\textrm{R}_{TM}=0$,  $\textrm{R}_{TM}=2$,  $\textrm{R}_{TM}=20$, respectively.}\label{fg5}
\end{figure}
number  $\textrm{Q}=1.68\cdot10^5$.  In Fig. \ref{fg5} we observe the decrease of the minimum critical Rayleigh number $\textrm{Ra}_{st}^{{min}}$  with an increase the coefficient  $\textrm{R}_{TM}$, which corresponds to a decrease  the threshold for the development of instability. Curve 1 is plotted for the case  $\textrm{R}_{TM}=0$, curve 2 --  $\textrm{R}_{TM}=2$, curve 3 -- $\textrm{R}_{TM}=20$. 

Thus, the conclusions about the lowering of the threshold of convective instability  taking into account TM effects remain valid even in the presence of an external magnetic field. 

\end{enumerate} 
	 
All the limiting cases considered above are completely in agreement with the conclusions of the works of Chandrasekhar \cite{42s} on the suppression of convection by the effects of rotation and an external magnetic field.

\begin{figure}
  \centering
	\includegraphics[width=8 cm, height=7 cm]{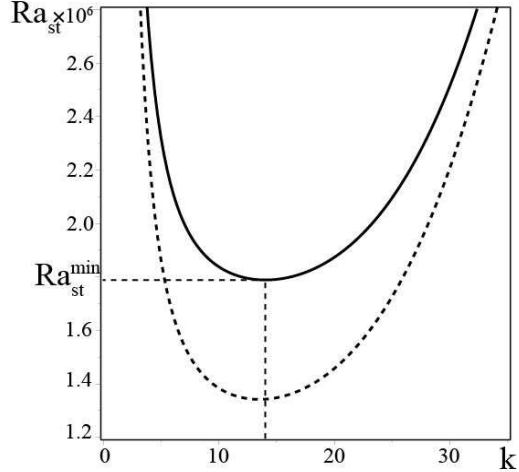}\\
	\caption{ Dependence of the stationary Rayleigh number $\textrm{Ra}_{st}$  on wavenumbers $k$ for a rotating medium (  $\textrm{Ta}=946$, $\textrm{Ro}=-1$ ) in an external magnetic field  $B_0=10^{-1}$T and the magnetic Prandtl number $\textrm{Pm}=1$. The solid line corresponds to the case without taking into account TM effects $q_\alpha=0, R_\alpha=0$, and the dashed line for $q_\alpha\neq 0, R_\alpha\neq 0$.}\label{fg6}
\end{figure}

Let us now proceed to study the general case of nonuniformly rotating stationary magnetoconvection taking into account TM effects. As before, we are calculated the convection parameters  $(\textrm{Q},\textrm{Ta},\textrm{Pm},q_\alpha,R_\alpha)$ using the values of physical quantities $(\rho_0, \nu, \eta, \chi, T_0, \alpha, B_0, \Omega_0 )$  for the Earth's core \cite{41s}: $\textrm{Q}=1.68\cdot10^5$, $\textrm{Ta}=946$,  $\textrm{Pm}=1$, $q_\alpha\textrm{Pm}\textrm{Pr}^{-1}\approx 5.24\cdot10^{-8}$, $\textrm{R}_\alpha\widetilde{\textrm{Q}}\approx2.5\cdot 10^9$.  Fig. \ref{fg6} shows the minimum value of the critical Rayleigh number  $\textrm{Ra}_{st}^{{min}}$ \ref{eq13m} for the case when there are no TM effects $\Delta\alpha=0$ (the medium is homogeneous in chemical composition). As seen from Fig. \ref{fg6}, when TM effects are taken into account the minimum critical Rayleigh number $\textrm{Ra}_{st}^{{min}}$  decreases, i.e. the threshold for the onset of convective instability decreases. The dependence plot  $\textrm{Ra}_{st}(k)$  (Fig. \ref{fg6}) is built for the Rayleigh rotation profile  $(\textrm{Ro}=-1)$. The dependence plot $\textrm{Ra}_{st}(k)$ for the profile of uniform rotation $(\textrm{Ro}=0)$ and positive profile $(\textrm{Ro}=2)$ has the similar view. It follows from the results obtained above that the generation of a magnetic field using TM effects promotes the development of convective instability. Magnetic and thermal perturbations are localized in convective cells on scales of 
the order $l \sim k_c^{-1}$.
       
\section{ Weakly nonlinear regime of convection taking into account thermomagnetic effects}

In this section, we will consider the weakly nonlinear convection regime, limiting ourselves, as in the previous section, to TM effects associated with the inhomogeneity of the thermoelectromotive force coefficient and "magnetization" of the heat flux (Leduc-Righi effect). By weakly nonlinear convection, we mean the interaction of small amplitudes of convective cells, which can be described as follows. Let the small amplitude of convective cells be of order  $O(\epsilon^1)$, then the interaction of the cells with each other leads to the second harmonic and nonlinearity of the order  $O(\epsilon^2)$, and then to nonlinearity  $O(\epsilon^3)$ , etc. In this case, the nonlinear terms in equations (\ref{eq9}) are considered as a perturbed response for the linear convection problem. In this case, the Rayleigh parameter  $\textrm{Ra}$ controlling convection is close to critical  $\textrm{Ra}_c$. Since the influence of 
\begin{figure}
  \centering
	\includegraphics[width=16 cm, height=7.0 cm]{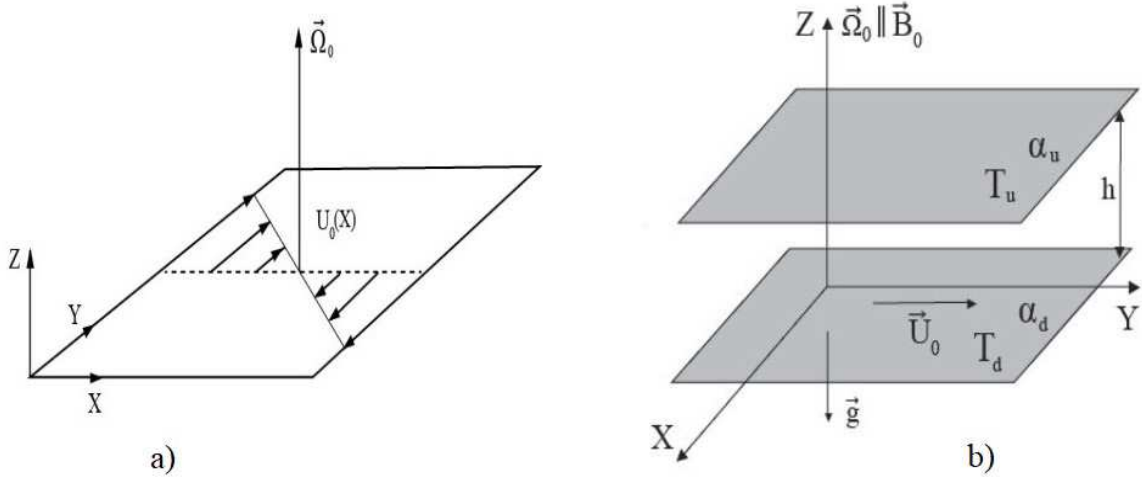} 
	\caption{(a) Diagram of a shear flow in rotating flows. In the local Cartesian system, this flow is approximated as a linear shear with velocity ${\bf{U}}_0(X)$; (b) Cartesian approximation of the problem for a nonuniformly rotating magnetic convection. Nonuniform rotation in the local Cartesian system of coordinates consists of rotation with constant angular velocity ${\bf{\Omega}}_0$ and shear velocity ${\bf{U}}_0$ directed along the Y-axis.}\label{fg7}
\end{figure}
unstable modes is small, our task is to obtain an equation that describes the interaction of these modes.

\subsection{ Equations of nonlinear convection in rotating flows of a magnetized fluid with a shear velocity}

To describe the nonlinear convective phenomena in the nonuniformly rotating layer of the electrically conducting fluid, it is convenient to turn from the cylindrical coordinate system   $(R,\phi,z)$  to the local Cartesian $(X,Y,Z)$  one. If we consider a fixed region of the fluid layer with a radius   $R_0$   and angular velocity of rotation  $\Omega_0=\Omega(R_0)$, then the coordinates  $X=R-R_0$  correspond to the radial direction, $Y=R_0(\phi-\phi_0)$  to azimuthal, and  $Z=z$  to vertical (see Fig. \ref{fg7}).
In this case, the fluid layer inhomogeneous rotation can be represented locally as the rotation with the constant angular velocity ${\bf{\Omega}}_0$ and azimuthal shear \cite{44s}, whose velocity profile is locally linear  ${\bf{U}}_0=-q {\Omega}_0 X {\bf{e}}_Y$, where $q\equiv -d \ln \Omega/d\ln R$  is the dimensionless shear parameter, determined from the profile of the angular velocity of rotation $\Omega (R)=\Omega_0 (R/R_0)^{-q}$. The shear parameter $q$ is bound up with the hydrodynamic Rossby number  $\textrm{Ro}=\frac{R}{2\Omega} \frac{\partial\Omega}{\partial R}$ by the relation $q=-2\textrm{Ro}$. Note, that the accretion disk with a shear parameter  $q=3/2, \textrm{Ro}=-3/4$ correspond to the Keplerian disk,  $q=2, \textrm{Ro}=-1$  corresponds to the disk with a constant angular momentum, or the Rayleigh rotation profile. The case of  $q=1, \textrm{Ro}=-1/2$ corresponds to the system with a flat rotation curve, and that of   $q=0, \textrm{Ro}=0$   to the homogeneous (or solid-state) rotation with a constant angular velocity. As before, we assume that the direction of the external magnetic field ${\bf B}_{0} $ coincides with the axis of rotation of the fluid  ${\bf \Omega }$ $\parallel OZ$.
    
The equations for the perturbations  $({\bf{u}}=(u_X,u_Y,u_Z), {\bf{b}}=(b_X,b_Y,b_Z), p, \theta)$   in the local Cartesian coordinate system take the following form:
\[\left(\frac{\partial}{\partial t}-\nu\nabla^2\right)u_X+({\bf {u}}\cdot\nabla)u_X-2\Omega_0u_Y=-\frac{1}{\rho_0}\frac{\partial \widetilde p}{\partial X}+\]
\begin{equation} \label{eq17m} 
+\frac{1}{\mu \rho_0}({\bf {b}}\cdot \nabla)b_X+\frac{B_{0}}{\mu\rho_0}\frac{\partial b_X}{\partial Z}
\end{equation}
 \[\left(\frac{\partial}{\partial t}-\nu\nabla^2\right)u_Y+({\bf {u}} \cdot\nabla)u_Y +2\Omega_0u_X(1+\textrm{Ro})=\]
\begin{equation} \label{eq18m}
=\frac{1}{\mu\rho_0}({\bf {b}} \cdot\nabla)b_Y+ \frac{B_{0}}{\mu\rho_0}\frac{ \partial b_Y }{\partial Z}
\end{equation}
\[\left(\frac{\partial}{\partial t}-\nu\nabla^2\right)u_Z+({\bf {u}}\cdot\nabla)u_Z=-\frac{1}{\rho_0}\frac{\partial \widetilde p}{\partial Z}+g\beta_T\theta+\]
\begin{equation} \label{eq19m} 
+\frac{1}{\mu\rho_0}({\bf {b}}\cdot \nabla)b_Z+\frac{B_{0}}{\mu\rho_0}\frac{\partial b_Z}{\partial Z}
\end{equation}
\begin{equation} \label{eq20m}
\left(\frac{\partial}{\partial t}-\eta\nabla^2\right)b_X-B_{0}\frac{\partial u_X}{\partial Z}+({\bf {u}} \cdot\nabla)b_X-({\bf {b}} \cdot\nabla)u_X=0
\end{equation}
\[\left(\frac{\partial}{\partial t}-\eta\nabla^2\right)b_Y-B_{0}\frac{\partial u_Y}{\partial Z}-2\Omega_0\textrm{Ro}b_X +({\bf {u}} \cdot\nabla)b_Y-\]
\begin{equation} \label{eq21m}
-({\bf {b}} \cdot\nabla)u_Y=\frac{\Delta\alpha}{h}\frac{\partial \theta}{\partial X}
\end{equation}
\begin{equation} \label{eq22m}
\left(\frac{\partial}{\partial t}-\eta\nabla^2\right)b_Z-B_{0}\frac{\partial u_Z}{\partial Z}+({\bf {u}}\cdot\nabla)b_Z-({\bf {b}}\cdot \nabla)u_Z=0
\end{equation}
\[\left(\frac{\partial}{\partial t}-\chi\nabla^2\right)\theta-u_Z\cdot \frac{\Delta T}{h} +({\bf {u}}\cdot\nabla)\theta=-\frac{\alpha T_0}{\rho_0c_p\mu h}\left(\frac{\Delta\alpha}{\alpha}+\frac{\Delta T}{T_0}\right)\frac{\partial b_Y}{\partial X}-\]
\[-\chi_{\wedge}\frac{\Delta T}{h}\frac{\partial b_Y}{\partial X}-
\frac{1}{\rho_0c_p\mu}\left(-\alpha\frac{\partial\theta}{\partial X}\frac{\partial b_Y}{\partial Z}+\alpha\frac{\partial\theta}{\partial Z}\frac{\partial b_Y}{\partial X}+\theta \frac{\Delta\alpha}{h}\frac{\partial b_Y}{\partial X}\right)-\]
\begin{equation} \label{eq23m}
-\chi_{\wedge}\left(\frac{\partial b_Y}{\partial X}\frac{\partial \theta}{\partial Z}-\frac{\partial b_Y}{\partial Z}\frac{\partial\theta}{\partial X}\right), 
\end{equation}
where the pressure  $\widetilde{p}$  also includes the perturbed magnetic pressure $p_m=\frac{1}{2\mu}\left(2{\bf B}_0\cdot{\bf b}+{\bf b}^2\right)$: $\widetilde{p}=p+p_m$. The action of nabla operators can be written as: 
$$({\bf a} \cdot \nabla)=a_x\cdot\frac{\partial}{\partial X}+a_z\cdot\frac{\partial}{\partial Z},\quad \nabla^2=\frac{\partial^2}{\partial X^2}+\frac{\partial^2}{\partial Z^2}. $$
In equations (\ref{eq17m})-(\ref{eq23m}), we assumed that all perturbed quantities depend only on two variables $(X,Z)$, i.e., we consider the dynamics of axisymmetric perturbations. To eliminate the pressure $\widetilde{p}$  in equations (\ref{eq17m}) and (\ref{eq19m}), we differentiate equation (\ref{eq17m}) according to  $Z$, and equation (\ref{eq19m}) according to  $X$, and then subtracting each other we obtain the equation for  $Y$-vortex component  $\textrm{rot}{\bf u}={\bf e}_Y \omega $:
 \[\left(\frac{\partial}{\partial t}-\nu\nabla^2\right)\omega+\frac{\partial}{\partial Z}\left(u_X \frac{\partial u_X}{\partial X}+u_Z\frac{\partial u_X}{\partial Z}\right)-\]
\[-\frac{\partial}{\partial X}\left(u_X\frac{\partial u_Z}{\partial X}+u_Z\frac{\partial u_Z}{\partial Z}\right)= \frac{B_{0}}{\mu\rho_0}\frac{\partial I}{\partial Z}+\frac{1}{\mu\rho_0}\times\]
\[\times \left(\frac{\partial}{\partial Z}\left(b_X \frac{\partial b_X}{\partial X}+ b_Z\frac{\partial b_X}{\partial Z}\right)-\frac{\partial}{\partial X}\left(b_X \frac{\partial b_Z}{\partial X}+b_Z \frac{\partial b_Z}{\partial Z}\right)\right) +\]
\begin{equation} \label{eq24m}
+ 2\Omega_0\frac{\partial  u_Y}{\partial Z}-g\beta_T \frac{\partial\theta}{\partial X},
\end{equation}
where  $\omega=\frac{\partial u_X}{\partial Z}-\frac{\partial u_Z}{\partial X}$  is the  $Y$-vortex component, $I=\frac{\partial b_X }{\partial Z}-\frac{\partial b_Z}{\partial X}$  is the $Y$-current component. The solenoidality equations for axisymmetric velocity and magnetic field perturbations take the form
\begin{equation} \label{eq25m} \frac{\partial u_X}{\partial X}+\frac{\partial u_Z}{\partial Z}=0, \quad \frac{\partial b_X}{\partial X}+\frac{\partial b_Z}{\partial Z}=0
\end{equation}	                                  	
Considering equations (\ref{eq25m}), we can introduce two scalar functions: hydrodynamic stream function $\psi$ and magnetic $\phi$, for which the following relations hold:
\[ u_X=-\frac{\partial\psi}{\partial Z},\quad u_Z=\frac{\partial\psi}{\partial X},\quad b_X=-\frac{\partial\phi}{\partial Z},\quad b_Z=\frac{\partial\phi}{\partial X}.  \]                                        
As a result, equations (\ref{eq24m}) and (\ref{eq18m}) take on a more compact form
\[ \left(\frac{\partial}{\partial t}-\nu\nabla^2\right)\nabla^2 \psi+2\Omega_0 \frac{{\partial u_Y}}{{\partial Z}} - \frac{{B_{0}}}{{\mu\rho _0 }}\frac{\partial }{{\partial Z}}\nabla ^2 \phi- g\beta_T \frac{{\partial \theta }}{{\partial X}} =\]
\begin{equation} \label{eq26m}
= \frac{1}{{\mu\rho_0 }}J(\phi,\nabla^2\phi )-J(\psi,\nabla^2 \psi ) \end{equation}
\[ \left(\frac{\partial}{\partial t}-\nu\nabla^2\right)u_Y- 2\Omega_0 (1+\textrm{Ro})\frac{{\partial \psi}}{{\partial Z}}-\frac{{B_{0}}}{{\mu\rho_0}}\frac{{\partial b_Y}}{{\partial Z}}=\]
\begin{equation} \label{eq27m} 
=\frac{1}{{\mu\rho_0 }}J(\phi,b_Y) - J(\psi,u_Y)
 \end{equation}             	
Here  $J(a,b) = \frac{{\partial a}}{{\partial X}}\frac{{\partial b}}{{\partial Z}} - \frac{{\partial a}}{{\partial Z}}\frac{{\partial b}}{{\partial X}}$   is the Jacobian or the Poisson bracket $J(a,b)\equiv \left\{a,b \right\}$.  
Further, differentiating equation (\ref{eq20m}) according to $Z$, and equation (\ref{eq22m}) according to $X$, and subtracting them from each other, we find the equation for $I$:
\[\left(\frac{\partial}{\partial t}-\eta\nabla^2\right)I+\frac{\partial}{\partial Z}\left(u_X\frac{\partial b_X}{\partial X}+u_Z\frac{\partial b_X}{\partial Z}- b_X \frac{\partial u_X}{\partial X}- b_Z\frac{\partial u_X}{\partial Z}\right)-\]
\begin{equation} \label{eq28m} 
-\frac{\partial}{\partial X}\left(u_X\frac{\partial b_Z}{\partial X}+u_Z\frac{\partial b_Z}{\partial Z}- b_X\frac{\partial u_Z}{\partial X}- b_Z \frac{\partial u_Z}{\partial Z}\right)=B_{0}\frac{\partial\omega}{\partial Z}
\end{equation}                                   
Equations (\ref{eq28m}) and (\ref{eq21m}) can also be written in a compact form using the definitions of stream functions $\psi$ and $\phi$:
\begin{equation} \label{eq29m}
\left(\frac{\partial}{\partial t}-\eta\nabla^2\right)\phi-B_{0}\frac{\partial\psi}{\partial Z}=-J(\psi,\phi)
\end{equation}
\[\left(\frac{\partial}{\partial t}-\eta\nabla^2\right)b_Y-B_{0}\frac{{\partial u_Y}}{{\partial Z}} + 2\Omega_0\textrm{Ro}\frac{{\partial \phi }}{{\partial Z}}-\frac{\Delta\alpha}{h}\frac{\partial \theta}{\partial X}=\]
\begin{equation} \label{eq30m}
=J(\phi,u_Y)-J(\psi,b_Y)\end{equation}                                     	
Similarly, we can write equation (\ref{eq23m}) for temperature perturbations using the definition of the hydrodynamic stream function $\psi$:
\[\left(\frac{\partial}{\partial t}-\chi\nabla^2\right)\theta-\frac{\Delta T}{h}\cdot\frac{\partial\psi}{\partial X}+\frac{\alpha T_0}{\rho_0c_p\mu h}\left[\left(\frac{\mu\mathcal{L}}{\alpha}+1\right)\frac{\Delta T}{T_0}+\frac{\Delta\alpha}{\alpha}\right]\frac{\partial b_Y}{\partial X}=\]
\begin{equation} \label{eq31m}
 =-J(\psi,\theta)-\frac{\Delta\alpha\cdot\theta}{\rho_0c_p\mu h}\frac{\partial b_Y}{\partial X}+\frac{\alpha}{\rho_0c_p\mu}\left(\frac{\mu\mathcal{L}}{\alpha}+1\right)J(\theta,b_Y)
\end{equation}
Equations (\ref{eq26m})-(\ref{eq27m}), (\ref{eq29m})-(\ref{eq31m}) are supplemented with the following boundary conditions:
\begin{equation} \label{eq32m}\psi|_{Z=0,h}=\nabla^2\psi|_{Z=0,h}=\frac{dv}{dZ}|_{Z=0,h}=\widetilde v|_{Z=0,h}=0, $$
$$ \frac{d \phi}{dZ}|_{Z=0,h}=\theta|_{Z=0,h}=0   \end{equation}                            	
To study the nonlinear convection regime, in equations (\ref{eq26m})-(\ref{eq27m}), (\ref{eq29m})-(\ref{eq31m}) it is convenient to pass to dimensionless variables:
\[ (X,Z)=h(x^*,z^*), \; t=\frac{h^2}{\nu}t^*,\; \psi=\chi\psi^*,\; \phi=hB_0 \phi^*,\] 
\[ u_Y=\frac{\chi}{h}v^*, \; b_Y=B_0 \widetilde v^*,\; \theta=(\Delta T)\theta^*. \]                                
Omitting the asterisk, we rewrite these equations in dimensionless variables:
 \[\left(\frac{\partial }{{\partial t}}- \nabla ^2 \right)\nabla^2\psi+\sqrt{\textrm{Ta}}\frac{{\partial v}}{{\partial z}}-\Pr\textrm{Pm}^{-1}\textrm{ Q}\frac{\partial}{{\partial z}}\nabla ^2 \phi-\textrm{Ra}\frac{{\partial \theta }}{{\partial x}}=\]
\[=\Pr\textrm{Pm}^{-1}\textrm{Q}\cdot J(\phi,\nabla^2 \phi )-{\Pr}^{-1}\cdot J(\psi,\nabla^2 \psi )\]
\[\left(\frac{{\partial }}{{\partial t}}-\nabla^2\right)v-\sqrt{\textrm{Ta}}(1 +\textrm{Ro})\frac{{\partial \psi}}{{\partial z}} - \Pr\textrm{Pm}^{ - 1}\textrm{Q}\frac{{\partial \widetilde v}}{{\partial z}}= \]
\[= \Pr\textrm{Pm}^{-1}\textrm{Q}\cdot J(\phi,\widetilde v)-{\Pr}^{-1}\cdot J(\psi,v) \]
\begin{equation} \label{eq33m}
 \left(\frac{{\partial}}{{\partial t}}-\textrm{Pm}^{-1}\nabla^2\right)\phi-{\Pr}^{-1}\frac{{\partial \psi }}{{\partial z}}=-{\Pr}^{-1}J(\psi,\phi )\end{equation}
\[\left(\frac{{\partial}}{{\partial t}}-\textrm{Pm}^{-1}\nabla^2 \right)\widetilde v-{\Pr}^{-1} \frac{{\partial v}}{{\partial z}} +\textrm{Ro}\sqrt{\textrm{Ta}}\frac{{\partial \phi }}{{\partial z}}-\textrm{Pm}^{-1}R_\alpha\frac{\partial\theta}{\partial x}=\]
\[={\Pr}^{-1}(J(\phi ,v)-J(\psi ,\widetilde v))\]
\[\left({\Pr}\frac{{\partial}}{{\partial t}}-\nabla^2\right)\theta-\frac{{\partial \psi}}{{\partial x}}+q_\alpha\frac{\partial \widetilde{v}}{\partial x}=-J(\psi,\theta)-q_\alpha^{(1)}\theta\cdot\frac{\partial \widetilde{v}}{\partial x}+q_\alpha^{(2)}J(\theta,\widetilde{v}),    \]        where new designations for dimensionless parameters are introduced
$$q_\alpha^{(1)}=\frac{\Delta\alpha B_0}{\rho_0c_p\mu \chi},\quad q_\alpha^{(2)}=\frac{\alpha B_0}{\rho_0c_p\mu \chi}\left(\frac{\mu\mathcal{L}}{\alpha}+1\right). $$
In the absence of thermal and thermomagnetic phenomena, the system of equations (\ref{eq33m}) was used to study the saturation mechanism of the standard MRI \cite{45s}. In the case when there are no TM effects, the system of equations (\ref{eq33m})  was used to study weakly nonlinear and chaotic convection regimes in a nonuniformly rotating plasma in an axial magnetic field \cite{39s}.

\subsection{Equation of finite amplitude for stationary convection}

We will obtain an equation for the finite amplitude of the magnetic field generated by the Rayleigh-Benard convection and thermomagnetic instability in a nonuniformly rotating electrically conductive fluid in an external uniform magnetic field using the weakly nonlinear theory (see for example \cite{46s}). We represent all the variables in equations (\ref{eq33m}) in the form of an asymptotic expansion:
 \[ \textrm{Ra}=\textrm{Ra}_c+\epsilon^2 \textrm{Ra}_2+\epsilon^4 \textrm{Ra}_4+\ldots \]
\[\psi=\epsilon \psi_1+\epsilon^2 \psi_2+\epsilon^3\psi_3+\ldots\]
\begin{equation} \label{eq1n}
 v=\epsilon v_1+\epsilon^2 v_2+\epsilon^3 v_3+\ldots  \end{equation}
\[\phi=\epsilon \phi_1+\epsilon^2 \phi_2+\epsilon^3\phi_3+\ldots\]
\[\widetilde v=\epsilon \widetilde v_1+\epsilon^2 \widetilde v_2+\epsilon^3\widetilde v_3+\ldots\]
\[\theta=\varepsilon \theta_1+\epsilon^2 \theta_2+\epsilon^3\theta_3+\ldots , \]                                       
where  $\epsilon$  is the small parameter of the expansion, which is the relative deviation of the Rayleigh number $\textrm{Ra}$ from the critical value $\textrm{Ra}_c$:\[\epsilon^2=\frac{\textrm{Ra}-\textrm{Ra}_c}{\textrm{Ra}_c} \ll 1 .\]
We assume that the amplitudes of the perturbed quantities depend only on the slow time  $\tau=\epsilon^2 t$.   Substituting expansions (\ref{eq1n}) into the system of equations (\ref{eq33m}), we solve it for different orders in   $\epsilon$.  For simplicity,  we will take into account the nonlinear terms in (\ref{eq33m}) only in the heat balance equation.

In the first order in  $\epsilon$, we obtain the equation 
   \begin{equation} \label{eq2n} \widehat{L}M_1=0, \end{equation}                                                	
where  \[ M_1=\left(\begin{array}{c}  \psi_1 \\ v_1 \\ \phi _1 \\ \widetilde v_1  \\ \theta _1  \end{array} \right),\]
$\widehat{L}$ is the matrix operator of the form:
\[\widehat L =\left(\begin{array}{ccccc}
    - \nabla^4  & \sqrt{\textrm{Ta}}\frac{\partial }{\partial z} & -\widetilde{\textrm{Q}}\frac{\partial}{\partial z}\nabla^2  & 0 & - \textrm{Ra}_c \frac{\partial }{\partial x}  \\ 
		
-\sqrt{\textrm{Ta}}(1+\textrm{Ro})\frac{\partial }{\partial z} & -\nabla^2  & 0 &  -\widetilde{\textrm{Q}}\frac{\partial}{\partial z} & 0  \\
- \Pr^{-1}\frac{\partial}{\partial z} & 0 &  -\textrm{Pm}^{-1} \nabla^2  & 0 & 0  \\

0 & - \Pr ^{-1}\frac{\partial}{\partial z} &  \textrm{Ro}\sqrt{\textrm{Ta}}\frac{\partial}{\partial z}  & -\textrm{Pm}^{-1}\nabla^2 & -\textrm{Pm}^{-1}R_\alpha\frac{\partial }{\partial x} \\
		
 -\frac{\partial}{\partial x} & 0 & 0 & q_\alpha\frac{\partial}{\partial x} & -\nabla^2   \\
\end{array} \right). \]
The solutions of the system of equations (\ref{eq2n}) with the boundary conditions of (\ref{eq32m}) have, respectively, the form:
\[\psi_1=A(\tau)\sin k_c x\sin\pi z,\quad \phi_1=\frac{A(\tau)\pi \textrm{Pm}}{a^2 \Pr}\sin k_c x\cos\pi z, \]
\[ \theta_1=\frac{A(\tau)k_c}{a^2}\left(1+q_\alpha\cdot\Pi_\alpha\right)\cos k_c x\sin\pi z, \quad  \widetilde v_1=-\Pi_\alpha A(\tau)\sin k_c x\sin\pi z, \]
\begin{equation} \label{eq3n} \end{equation}
\[ v_1=\frac{\pi\sqrt{\textrm{Ta}}[(1+\textrm{Ro})(a^4-k_c^2q_\alpha R_\alpha)+\pi^2\textrm{QPmRo}]-\pi k_c^2\widetilde{\textrm{Q}} R_\alpha}{a^2(a^4+\pi^2 \textrm{Q}-k_c^2q_\alpha R_\alpha)} \times\]
\[\times A(\tau)\sin k_c x \cos\pi z, \]
where 
\[\Pi_\alpha=\frac{\pi^2\textrm{Pm}\Pr ^{-1}\sqrt{\textrm{Ta}}(1+\textrm{Ro}(1-\textrm{Pm}))+k_c^2 R_\alpha}{a^4+\pi^2\textrm{Q}-k_c^2q_\alpha R_\alpha} .\] 
The critical value of the Rayleigh number $\textrm{Ra}_c$  for stationary convection is found from the first equation of system (\ref{eq2n}) and has the form of the formula (\ref{eq13m}) obtained in linear theory. The amplitude $A(\tau)$ is still unknown.

For the second-order in  $\epsilon$, we have the following equation:
\begin{equation} \label{eq4n} \widehat{L}M_2=N_2, \end{equation}
where $M_2=\left(\begin{array}{c} \psi _2 \\ v_2 \\ \phi _2 \\ \widetilde v_2  \\ \theta _2 \end{array} \right)$,\quad $N_2=\left(\begin{array}{c} N_{21} \\ N_{22} \\ N_{23} \\ N_{24}  \\ N_{25} \end{array} \right)$, 
\[N_{21}=N_{22}=N_{23}=N_{24}=0,\]
\[N_{25}=-\left(\frac{\partial\psi_1}{\partial x}\frac{\partial \theta_1}{\partial z}-\frac{\partial \theta_1}{\partial x}\frac{\partial\psi_1}{\partial z}\right)-q_\alpha^{(1)}\theta_1 \frac{\partial \widetilde v_1}{\partial x} +q_\alpha^{(2)}\left(\frac{\partial\theta_1}{\partial x}\frac{\partial \widetilde v_1}{\partial z}-\frac{\partial \theta_1}{\partial z}\frac{\partial\widetilde v_1}{\partial x}\right)  .\]
Using solutions (\ref{eq3n}) and boundary conditions (\ref{eq32m}), we can find solutions of equations (\ref{eq4n}):
\[\psi_2=0,\quad \phi_2=0, \quad v_2=0,\quad \widetilde v_2=0 , \]
 \begin{equation} \label{eq5n}
\theta_2=-\frac{A^2(\tau)k_c^2}{8\pi a^2}\left(1-q_\alpha^{(2)}\cdot\Pi_\alpha\right)\left(1+q_\alpha\cdot\Pi_\alpha\right)\sin(2\pi z).   
\end{equation}                              	
To analyze the intensity of the heat transfer, horizontally-averaged heat flux is introduced at the boundary of the layer of electrically conducting fluid (Nusselt number)
 \begin{equation} \label{eq6n}
\textrm{Nu}(\tau)=1+\frac{\left[\frac{k_c}{2\pi}\int\limits_0^{2\pi/k_c}\left(\frac{\partial \theta_2}{\partial z}\right)dx\right]_{z=0}}{\left[\frac{k_c}{2\pi}\int\limits_0^{2\pi/k_c}\left(\frac{\partial T_0}{\partial z}\right)dx\right]_{z=0}}=1+\frac{k_c^2}{4a^2}A^2(\tau)
\end{equation}
The heat flow intensity (of Nusselt number $\textrm{Nu}$) will be analyzed after the expression for the amplitude  $A(\tau)$  is obtained.

For the third order of $\epsilon$  we can find:
\begin{equation} \label{eq7n} \widehat{L}M_3=N_3, \end{equation}
where $M_3=\left(\begin{array}{c} \psi _3 \\ v_3 \\ \phi _3 \\ \widetilde v_3  \\ \theta _3  \end{array} \right)$,\quad  $N_3=\left(\begin{array}{c} N_{31} \\ N_{32} \\ N_{33} \\ N_{34}  \\ N_{35} \end{array} \right)$; 
\[N_{31}=-\frac{\partial}{\partial\tau}\nabla^2\psi_1+\textrm{Ra}_2 \frac{\partial\theta_1}{\partial x}=\left(a^2\frac{\partial A(\tau)}{\partial \tau}-\textrm{Ra}_2 \frac{k_c^2 A(\tau)}{a^2}\left(1+q_\alpha\cdot\Pi_\alpha\right)\right)\times\]
\[\times\sin k_c x \sin\pi z,   \]

\[N_{32}=-\frac{\partial v_1}{\partial\tau}=-\frac{\pi\sqrt{\textrm{Ta}}[(1+\textrm{Ro})(a^4-k_c^2q_\alpha R_\alpha)+\pi^2\textrm{QPmRo}]-\pi k_c^2\widetilde{\textrm{Q}}R_\alpha}{a^2(a^4+\pi^2\textrm{Q}-k_c^2q_\alpha R_\alpha)}\times \]
\[\times\frac{\partial A(\tau)}{\partial \tau}\sin k_c x \cos \pi z, \]

\[N_{33}=-\frac{\partial\phi_1}{\partial\tau}=-\frac{\pi \textrm{Pm}}{a^2 \Pr}\cdot\frac{\partial A(\tau)}{\partial \tau}\sin k_c x \cos \pi z    \]

\[N_{34}=-\frac{\partial \widetilde v_1}{\partial\tau}= \Pi_\alpha\cdot\frac{\partial A(\tau)}{\partial \tau}\sin k_c x \sin \pi z , \]

\[N_{35}=-\Pr\frac{\partial\theta_1}{\partial\tau}-\frac{\partial\psi_1}{\partial x}\frac{\partial\theta_2}{\partial z}-q_\alpha^{(1)}\theta_2\frac{\partial \widetilde{v}_1}{\partial x}-q_\alpha^{(2)}\frac{\partial\theta_2}{\partial z}\frac{\partial\widetilde{v}_1}{\partial x}=\]
\[ =-\Pr \frac{k_c}{a^2}\frac{\partial A(\tau)}{\partial \tau}\cos k_c x\sin \pi z+\frac{k_c^3 A^3(\tau)}{4a^2}(1-q_\alpha^{(2)}\cdot \Pi_\alpha)^2(1+q_\alpha \cdot \Pi_\alpha)\times\]
\[\times\cos k_c x \sin \pi z\cos 2 \pi z-\]
\[-q_\alpha^{(1)}\cdot\frac{k_c^3A^3(\tau)}{8\pi a^2}(1-q_\alpha^{(2)}\cdot \Pi_\alpha)(1+q_\alpha \cdot \Pi_\alpha)\Pi_\alpha\cos k_c x \sin \pi z\sin 2\pi z . \]
The solvability condition for the chain of nonlinear equations (\ref{eq4n}), (\ref{eq7n}) is known as Fredholm's alternative (see, for example, \cite{47s})
 \begin{equation} \label{eq8n} 
 \left\langle M_1^{\dagger}, R.H.\right\rangle =0,
 \end{equation}
where  $R.H.$ are the right sides of the perturbed equations with nonlinear terms. The matrix  $M_1^{\dagger} =(\psi_1^{\dagger},\theta_1^{\dagger},\phi_1^{\dagger},v_1^{\dagger})^{Tr}$ is a nontrivial solution to a linear self-adjoint problem  $ \widehat{L}^{\dagger} M_1^{\dagger}= 0 $, where $ \widehat{L}^{\dagger} $ is a self-adjoint operator, which is determined from the following relation
 \begin{equation} \label{eq9n} 
 \left\langle M_1^{\dagger},\widehat{L} M_1 \right \rangle \equiv \left\langle   \widehat{L}^{\dagger}M_1^{\dagger}, M_1 \right\rangle,  \end{equation}                           	
where $\left\langle , \right\rangle$  is the inner product, which here has the following definition:
  $$\left\langle {\bf{f}} , {\bf{g}}\right\rangle= \int\limits_{z=0}^1\int\limits_{x=0}^{2\pi/k_c} {\bf{f}}\cdot {\bf{g}}\, dxdz .  $$         
Using expression (66), we write the Fredholm solvability condition for third-order $O(\varepsilon^3)$  equations (\ref{eq7n}) in the following form:
 \[\int\limits_{z=0}^1\int\limits_{x=0}^{2\pi/k_c}\left[\widehat{\mathcal K}\widehat{\mathcal M} \psi_1^{\dagger}\cdot {\mathcal R}_{31}-\textrm{Ra}_c \frac{\partial}{\partial x}\widehat{\mathcal K} \theta_1^{\dagger}\cdot {\mathcal R}_{32}+\widetilde{\textrm{Q}}\nabla^2 \widehat{\mathcal K}\widehat{\mathcal M}\phi_1^{\dagger}\cdot {\mathcal R}_{33}+\right. \]
\begin{equation} \label{eq10n}
\left. +\sqrt{\textrm{Ta}}\frac{\partial}{\partial z}\widehat{\mathcal M} v_1^{\dagger}\cdot {\mathcal R}_{34}\right]dxdz=0,
\end{equation}
where the notations are introduced
\[ \widehat{\mathcal K}=\sqrt{\textrm{Ta}}\widehat P\frac{\partial}{\partial z}\left(\nabla^4+q_\alpha R_\alpha\frac{\partial^2}{\partial x^2}\right)+\widetilde{\textrm{Q}}R_\alpha\frac{\partial^3 \widehat{q}}{\partial x^2\partial z}, \]
\[\widehat{\mathcal M}=-\frac{\partial^2 \widehat{q}}{\partial x\partial z}\left(\nabla^4-\textrm{Q}\frac{\partial^2}{\partial z^2}\right)+q_\alpha\textrm{Pm}\textrm{Pr}^{-1}\sqrt{\textrm{Ta}}\widehat P\frac{\partial^4}{\partial x\partial z^3}, \]
\[\widehat P=(1+\textrm{Ro})\nabla^4-\textrm{QPmRo}\frac{\partial^2}{\partial z^2},\; \widehat{q}=\nabla^4+q_\alpha\textrm{Pm}^2\textrm{Pr}^{-1}\textrm{Ro}\sqrt{\textrm{Ta}}\frac{\partial^2}{\partial z^2}, \]
\[ {\mathcal R}_{31}=N_{31},\quad {\mathcal R}_{32}= \left(\nabla^4-\textrm{Q}\frac{\partial^2}{\partial z^2}\right)\textrm{R}_{32}+q_\alpha\textrm{Pm}\textrm{Pr}^{-1}\frac{\partial^3 \textrm{R}_{34}}{\partial x\partial z^2} , \]
\[{\mathcal R}_{33}=N_{33}\Pr, \quad {\mathcal R}_{34}=\left(\nabla^4+q_\alpha R_\alpha\frac{\partial^2}{\partial x^2}\right)\textrm{R}_{34}-\widetilde{\textrm{Q}}R_\alpha\frac{\partial\textrm{R}_{32}}{\partial x}, \]
\[ \textrm{R}_{32} = \frac{\partial \nabla^2}{\partial z}\left(\nabla^2N_{35}+q_\alpha\textrm{Pm}\frac{\partial N_{34}}{\partial x}\right)+q_\alpha\textrm{Pm}^2\textrm{Pr}^{-1}\textrm{Ro}\sqrt{\textrm{Ta}}\frac{\partial^3N_{33}}{\partial x\partial z^2}, \] 
\[ \textrm{R}_{34}=-\nabla^4 N_{32}+\textrm{QPr}\frac{\partial}{\partial z}\nabla^2 N_{34}+\sqrt{\textrm{Ta}}\textrm{QPmPrRo}\frac{\partial^2 N_{33}}{\partial z^2} . \]
Expressions for $\psi_1^{\dagger},\theta_1^{\dagger},\phi_1^{\dagger},v_1^{\dagger}$  are determined from the solution of the linear self-adjoint problem  $\widehat L^{\dagger} M_1^{\dagger}=0$: 
\[\psi_1^{\dagger}=A(\tau)\sin k_c x \sin \pi z,\]
\[ \theta_1^{\dagger}=-\frac{A(\tau)k_c}{a^2}\left(1+q_\alpha\cdot\Pi_\alpha\right)\cos k_c x\sin\pi z,   \]
\[ \phi_1^{\dagger}=-\frac{A(\tau)\pi \textrm{Pm}}{a^2 \Pr}\sin k_c x \cos \pi z ,\]
\[v_1^{\dagger}=-\frac{\pi\sqrt{\textrm{Ta}}[(1+\textrm{Ro})(a^4-k_c^2q_\alpha R_\alpha)+\pi^2\textrm{QPmRo}]-\pi k_c^2\widetilde{\textrm{Q}} R_\alpha}{a^2(a^4+\pi^2 \textrm{Q}-k_c^2q_\alpha R_\alpha)}\cdot A(\tau)\sin k_c x \cos\pi z .  \] 
  \begin{figure}
  \centering
	\includegraphics[width=13 cm, height=7 cm]{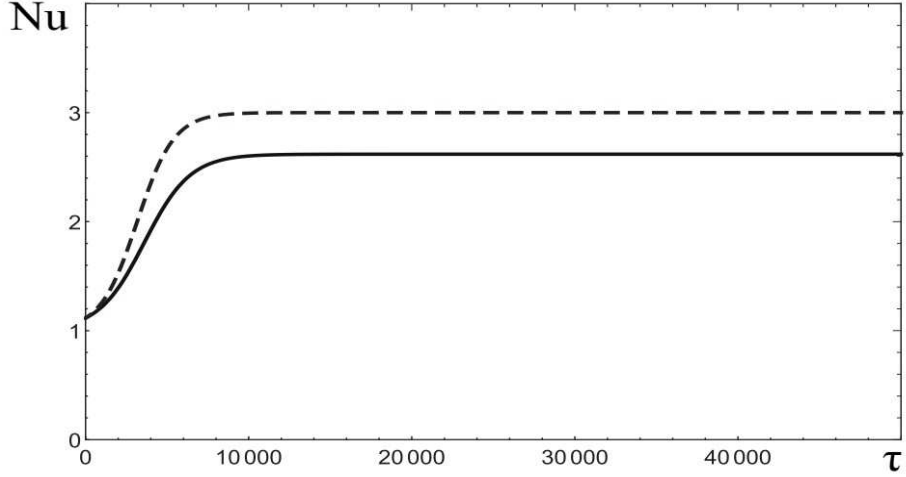}\\
	\caption{Dependence of the Nusselt number $\textrm{Nu}$ on time  $\tau$. The dashed line shows the case without taking into account TM effects and the solid line with taking into account TM effects.}\label{fg8}
\end{figure}                                                  
The self-adjoint matrix operator  $\widehat L^{\dagger}$   takes the following form:
 \[	\widehat L^{\dagger}=\left(\begin{array}{cccc}
 \nabla ^4\widehat{\mathcal K}\widehat{\mathcal M}   & -\textrm{Ra}_c\frac{\partial}{\partial x}\widehat{\mathcal K}\widehat{\mathcal M} &   -\widetilde{\textrm{Q}}\frac{\partial }{\partial z}\nabla^2\widehat{\mathcal K}\widehat{\mathcal M} & \sqrt {\textrm{Ta}}\frac{\partial }{\partial z}\widehat{\mathcal K}\widehat{\mathcal M}  \\ 
		 
-\textrm{Ra}_c\frac{\partial }{\partial x}\widehat{\mathcal K}\widehat{\mathcal M}  &  \textrm{Ra}_c\frac{\partial}{\partial x}\widehat{\mathcal K}\widehat{\mathcal N}  & 0 & 0  \\
		
-\widetilde{\textrm{Q}}\frac{\partial }{\partial z}\nabla^2\widehat{\mathcal K}\widehat{\mathcal M} & 0 & \widetilde{ \textrm{Q}}\Pr\textrm{Pm}^{-1}\nabla^4\widehat{\mathcal K}\widehat{\mathcal M}  & 0   \\

 \sqrt {\textrm{Ta}}\frac{\partial }{\partial z}\widehat{\mathcal K}\widehat{\mathcal M}  & 0 &  0  &  -\sqrt {\textrm{Ta}}\frac{\partial }{\partial z}\widehat{\mathcal L}\widehat{\mathcal M}\\
	\end{array} \right) \]
  Here the notations for new operators are inroduced
\[\widehat{\mathcal L}=q_\alpha R_\alpha\textrm{Q}\frac{\partial^4\nabla^2}{\partial x^2\partial z^2} +\nabla^2\left(\nabla^4-\textrm{Q}\frac{\partial^2}{\partial z^2}\right)\left(\nabla^4+q_\alpha R_\alpha\frac{\partial^2}{\partial x^2}\right),\]
\[\widehat{\mathcal N}=-\frac{\partial\nabla^2}{\partial z}\left(\nabla^4-\textrm{Q}\frac{\partial^2}{\partial z^2}\right)\left(\nabla^4+q_\alpha R_\alpha\frac{\partial^2}{\partial x^2}\right)-q_\alpha R_\alpha\textrm{Q}\frac{\partial^5\nabla^2}{\partial x^2\partial z^3}.\]                      
By integrating  (\ref{eq10n}), we can obtain a nonlinear equation for the amplitude  $A(\tau)$, which refers to the Ginzburg-Landau equation or the Bernoulli differential equation with constant coefficients:
\begin{equation} \label{eq12n} 
\mathscr A_1\frac{\partial A}{\partial\tau}-\mathscr A_2 A+\mathscr A_3 A^3=0
 \end{equation}
here  $\mathscr A_{1,2,3}$ are constant coefficients.
Because of their cumbersome form, expressions $\mathscr A_{1,2,3}$ are given in Appendix ${\rm B}$. In the limiting case, when TM effects are absent  $(q_\alpha=0, R_\alpha=0)$,  the equation (\ref{eq12n}) corresponds to the well-known result \cite{48s}. It is easy to obtain an analytical solution (\ref{eq12n}) with a known initial condition $A_0=A(0)$: 
 	\begin{equation} \label{eq13n}   A(\tau)=\frac{A_0}{\sqrt{\frac{\mathscr A_3}{\mathscr A_2}A_0^2+\left(1-A_0^2\frac{\mathscr A_3}{\mathscr A_2}\right)\exp\left(-\frac{2\tau\mathscr A_2}{\mathscr A_1}\right)}}
 \end{equation}
With the help of solution (\ref{eq13n}), we can determine the change of the magnitude of heat transfer (Nusselt number $\textrm{Nu}$ ) and the amplitude of the generated magnetic field $\widetilde{v}(\tau)$ 
\begin{figure}
  \centering
	\includegraphics[width=13 cm, height=7 cm]{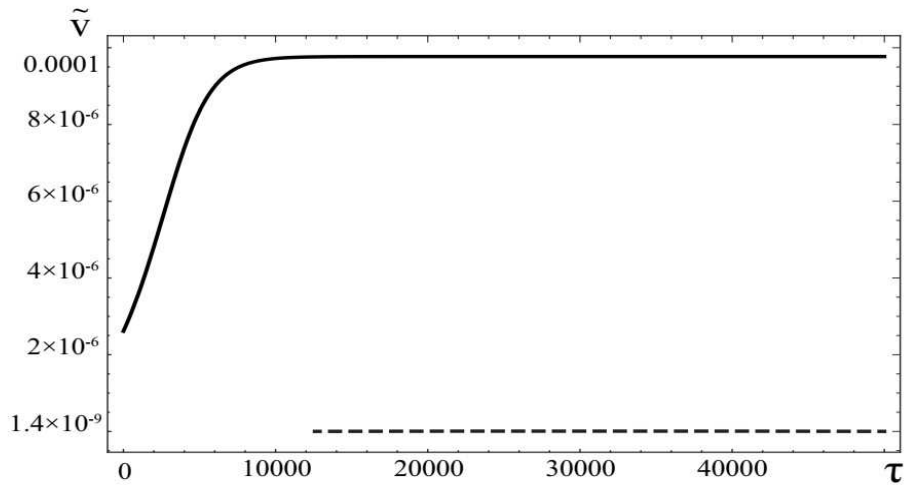}\\
	\caption{Dependence of the generated magnetic field amplitude  $\widetilde{v}(\tau)$  on time $\tau$. The steady-state amplitude $\widetilde{v}(\tau)$ taking into account TM effects (solid line) is approximately  $10^4$ times greater than for the case without TM effects (dashed line).}\label{fg9}
\end{figure}
from time $\tau$ using formula (\ref{eq6n}). When performing calculations, we take the initial amplitude equal to $A_0=0.7$  and $\textrm{Ra}_2\approx \textrm{Ra}_c$, which corresponds to the smallness of the supercriticality parameter  $\epsilon$. Constant convection parameters  $\textrm{Q}_1=\textrm{Q}/\pi^2 =17000$,  $\textrm{T}_1=\textrm{Ta}/\pi^4=10$,  $\textrm{R}_1=\textrm{Ra}/\pi^4=20000$, $\textrm{Pm}=1$, $\textrm{Pr}=380000$ correspond to the previously adopted parameters of the Earth's core and the external magnetic field $B_0=10^{-1}$T. The profile of nonuniform rotation (Rossby number) is assumed to be Rayleigh, i.e. $\textrm{Ro}=-1$. The graph of the dependence of the Nusselt number $\textrm{Nu}(\tau)$ for the above parameters is shown in Fig. \ref{fg8}. Here, the dashed line corresponds to the case without taking into account TM effects, and the solid line - with taking into account TM effects. The graphs show the establishment of the final value $\textrm{Nu}(\tau)$, due to the relationship between the number $\textrm{Nu}(\tau)$  and the amplitude $A(\tau)$  (see formula (\ref{eq6n})). The excess of the number $\textrm{Nu}$  over one is due to the occurrence of convection. When TM effects are taken into account, heat transfer due to convection decreases, since convective instability reaches a stationary level at a lower value of the final amplitude. In this case,  the part of the thermal energy is transformed into the energy of the generated magnetic field. The graph in Fig. \ref{fg9} shows the establishment of a finite amplitude for the generated disturbances of the toroidal ($Y$-component) magnetic field. Hence, it can be seen that when the TM effects are taken into account, the amplitude of the exciting magnetic field increased by about a factor of  $10^4$ one! 
      
Thus, convective processes taking into account TM phenomena play an essential role in the generation of magnetic fields in a nonuniformly rotating electrically conducting fluid.

\section{Conclusion}

In this paper, we consider the mechanism of magnetic field generation in a nonuniformly rotating electrically conductive fluid by TMI, which occurs at collinear temperature $\nabla T_0$ and thermoelectromotive force coefficient $\nabla\alpha $ gradients: $[\nabla\alpha\times\nabla T_0]=0$.  The gradient of the thermoelectromotive force coefficient $\nabla\alpha $ is caused by the inhomogeneity of the chemical composition of the electrically conductive fluid. We have investigated the generation of a magnetic field by TMI in a nonuniformly rotating layer of an electrically conductive fluid in constant vertical magnetic ${\bf{B}}_0 \| OZ$ and gravitational  $-{\bf{g}} \| OZ$   fields. In the linear approximation, we obtained the dispersion equation for axisymmetric perturbations, from which the critical Rayleigh number $\textrm {Ra}_c$  was determined for the stationary convection regime. The performed analysis of the stationary convection regime showed that the threshold for the development of convective instability decreases for a negative rotation profile ($\textrm{Ro}<0$). Also, the threshold for the development of convective instability   taking into account TM effects also  decreases  for any profile of a nonuniform rotation, i.e. has a destabilizing effect. We investigated the weakly nonlinear stage of stationary convection  taking into account  TM effects using perturbation theory in the small supercriticality parameter  $ \epsilon = \sqrt {(\textrm {Ra}-\textrm {Ra}_c) / \textrm {Ra}_c} $ of the stationary Rayleigh number  $\textrm {Ra}_c$  and obtained the nonlinear Ginzburg-Landau equation for the convection amplitude. From the solution of this equation, it follows that the generated toroidal magnetic field reaches a stationary level. 

The results obtained in this work can find application in various problems of the magnetic geodynamo, as well as in laboratory studies on rotating magnetic convection taking into account thermomagnetic phenomena.

\section*{Appendix}

\section*{ A. Differential equation operators (\ref{eq9m}) }

\[\widehat a_{11}=\widehat L_\nu-\frac{{\textrm{Q}\widehat D^2 }}{\widehat L_\eta +R_{\mathcal{N}}\widehat D}-\frac{ik\textrm{Ra}\widehat D}{\widehat D^2-k^2}\cdot\frac{\widehat{a}_2\widehat{m}}{\widehat{m}\widehat{l}-\widehat{a}_3\widehat{b}_2}+ \]
\[ +\frac{{\widetilde{\textrm{Q}}\widehat D }}{\widehat L_\eta +R_{\mathcal{N}}\widehat D}\left[ikR_{\mathcal{N}}\widehat D\cdot\frac{\widehat{a}_2\widehat{b}_2}{\widehat{m}\widehat{l}-\widehat{a}_3\widehat{b}_2}+R_{\mathcal{H}}\widehat D^2\cdot\frac{\widehat{a}_2\widehat{m}}{\widehat{m}\widehat{l}-\widehat{a}_3\widehat{b}_2}\right],\]
\[\widehat a_{12}=\frac{{\sqrt{\textrm{Ta}}\widehat D^2 }}{{\widehat D^2-k^2 }}-\frac{ik\textrm{Ra}\widehat D}{\widehat D^2-k^2}\cdot\frac{\widehat{a}_1\widehat{b}_2}{\widehat{m}\widehat{l}-\widehat{a}_3\widehat{b}_2}-\]
\[-\frac{{\widetilde{\textrm{Q}}\widehat D }}{\widehat L_\eta +R_{\mathcal{N}}\widehat D}\left[ikR_{\mathcal{N}}\widehat D\cdot\frac{\widehat{a}_1\widehat{b}_2}{\widehat{m}\widehat{l}-\widehat{a}_3\widehat{b}_2}+R_{\mathcal{H}}\widehat D^2\cdot\frac{\widehat{a}_1\widehat{m}}{\widehat{m}\widehat{l}-\widehat{a}_3\widehat{b}_2}\right],\]
\[\widehat a_{12}=\frac{{\widetilde{\textrm{Q}}}}{\widehat L_\eta +R_{\mathcal{N}}\widehat D}\left[\frac{(ikR_{\mathcal{N}}\widehat l+R_{\mathcal{H}}\widehat D\widehat{a}_3)\widehat{D}\widehat{b}_1}{\widehat{m}\widehat{l}-\widehat{a}_3\widehat{b}_2}\right]+\frac{ik\textrm{Ra}\widehat D}{\widehat D^2-k^2}\cdot\frac{\widehat{b}_1\widehat{l}}{\widehat{m}\widehat{l}-\widehat{a}_3\widehat{b}_2},\]
\[\widehat a_{21}=-\sqrt{\textrm{Ta}}(1+\textrm{Ro})+\widetilde{\textrm{Q}}\widehat D\cdot\frac{\widehat{a}_2\widehat{m}}{\widehat{m}\widehat{l}-\widehat{a}_3\widehat{b}_2}, \quad \widehat a_{22}=\widehat L_\nu-\widetilde{\textrm{Q}}\widehat D\cdot\frac{\widehat{a}_1\widehat{m}}{\widehat{m}\widehat{l}-\widehat{a}_3\widehat{b}_2},  \] 
\[\widehat a_{23}=\widetilde{\textrm{Q}}\widehat D\cdot\frac{\widehat{a}_3\widehat{b}_1}{\widehat{m}\widehat{l}-\widehat{a}_3\widehat{b}_2},\; \widehat a_{31}=-\frac{{ikR_{\mathcal{N}}\widetilde{\textrm{Q}}\widehat D^2}}{\widehat L_\eta+R_{\mathcal{N}}\widehat D}\cdot\left[\frac{\widehat{a}_2}{\widehat{l}}+\frac{\widehat{a}_2\widehat{a}_3\widehat{b}_2}{\widehat{l}(\widehat{m}\widehat{l}-\widehat{a}_3\widehat{b}_2)}\right]+ \]
\[+\frac{\textrm{Ra}\widehat D^2}{\widehat D^2-k^2}\cdot\frac{\widehat{a}_2\widehat{b}_2}{\widehat{m}\widehat{l}-\widehat{a}_3\widehat{b}_2}-\left[\textrm{Ra}-R_{\mathcal{N}}\frac{k^2\widetilde{\textrm{Q}}\widehat D}{\widehat L_\eta+R_{\mathcal{N}}\widehat D}\right]\frac{\widehat{a}_2\widehat{b}_2}{\widehat{m}\widehat{l}-\widehat{a}_3\widehat{b}_2}, \]  
\[\widehat a_{32}=\frac{{ikR_{\mathcal{H}}\widetilde{\textrm{Q}}\widehat D^2}}{\widehat L_\eta+R_{\mathcal{N}}\widehat D}\cdot\frac{\widehat{a}_1\widehat{m}}{\widehat{m}\widehat{l}-\widehat{a}_3\widehat{b}_2}-\frac{{ik\sqrt{\textrm{Ta}}\widehat D}}{{\widehat D^2-k^2}}-\]
\[ -\left[\frac{{k^2R_{\mathcal{N}}\widetilde{\textrm{Q}}\widehat D}}{\widehat L_\eta+R_{\mathcal{N}}\widehat D}+\frac{k^2\textrm{Ra}\widehat D}{\widehat D^2-k^2}\right]\cdot\frac{\widehat{a}_1\widehat{b}_2}{\widehat{m}\widehat{l}-\widehat{a}_3\widehat{b}_2}, \]
\[\widehat a_{33}=\widehat L_\nu-\frac{\textrm{Q}\widehat D^2}{\widehat L_\eta+R_{\mathcal{N}}}+\left[\frac{{k^2R_{\mathcal{N}}\widetilde{\textrm{Q}}\widehat D}}{\widehat L_\eta+R_{\mathcal{N}}\widehat D}+\frac{k^2\textrm{Ra}}{\widehat D^2-k^2}\right]\cdot\frac{\widehat{b}_1\widehat{l}}{\widehat{m}\widehat{l}-\widehat{a}_3\widehat{b}_2}-\]
\[-\frac{{ikR_{\mathcal{H}}\widetilde{\textrm{Q}}\widehat D^2}}{\widehat L_\eta+R_{\mathcal{N}}\widehat D}\cdot\frac{\widehat{a}_3\widehat{b}_1}{\widehat{m}\widehat{l}-\widehat{a}_3\widehat{b}_2},\]
\[\widehat{a}_1=\textrm{Pm}\textrm{Pr}^{-1}\widehat D(\widehat L_\eta+R_{\mathcal{N}}\widehat D),\quad \widehat{a}_2=\textrm{Pm}\textrm{Pr}^{-1}(\textrm{PmRo}\sqrt{\textrm{Ta}}-R_{\mathcal{H}}(\widehat D^2-k^2))\widehat D, \]
\[\widehat{a}_3=ik\left(R_\alpha(\widehat L_\eta+R_{\mathcal{N}}\widehat D)-R_{\mathcal{N}}(\textrm{PmRo}\sqrt{\textrm{Ta}}-R_{\mathcal{H}}(\widehat D^2-k^2))\widehat D\right),\]
\[ \widehat{b}_1= 1+\frac{q_{\mathcal{N}}\textrm{Pm}\textrm{Pr}^{-1}\widehat D(\widehat D^2-k^2)}{\widehat L_\eta+R_{\mathcal{N}}\widehat D},\quad \widehat{b}_1= \frac{ikR_{\mathcal{H}}\widehat D(\widehat D^2-k^2)}{\widehat L_\eta+R_{\mathcal{N}}\widehat D}-ikq_\alpha, \]
\[\widehat{m}= \widehat L_\chi+\frac{q_{\mathcal{N}}R_{\mathcal{N}}k^2(\widehat D^2-k^2)}{\widehat L_\eta+R_{\mathcal{N}}\widehat D}, \; \widehat{l}=\left(\widehat L_\eta+R_{\mathcal{N}}\widehat D\right)^2+\]
\[+R_{\mathcal{H}}\left(\textrm{PmRo}\sqrt{\textrm{Ta}}-R_{\mathcal{H}}(\widehat D^2-k^2)\right)\widehat D^2.\]

\section*{ B. Coefficients in the Ginzburg-Landau equation (\ref{eq12n}) }

\[ \mathscr A_1=\frac{a^2}{\Pr}+\frac{k_c^2\textrm{Ra}_c}{a^4}\times\]
\[\times\frac{\left(1+q_\alpha\Pi_\alpha\left(1+\frac{\textrm{Pm}}{\textrm{Pr}}\right)-q_\alpha\frac{\pi^2\textrm{Pm}^3}{a^4 \textrm{Pr}^2}\textrm{Ro}\sqrt{\textrm{Ta}}\right)\left(a^4+\pi^2\textrm{Q}\right)+ q_\alpha\frac{\pi\textrm{Pm}}{a^2 \textrm{Pr}^2}\cdot\delta }{a^4+\pi^2\textrm{Q}-q_\alpha R_\alpha k_c^2}-\]
\[- \frac{\pi^2\textrm{QPm}}{a^2\Pr}-\frac{\pi\sqrt{\textrm{Ta}}}{a^6(a^4+\pi^2\textrm{Q}-k_c^2q_\alpha R_\alpha)\textrm{Pr}}\times \]
\[ \times\left[\delta\cdot(a^4-k_c^2q_\alpha R_\alpha)-\pi k_c^2a^2\widetilde{\textrm{Q}}\textrm{Pr}R_\alpha\left(1+q_\alpha\Pi_\alpha\left(1+\frac{\textrm{Pm}}{\textrm{Pr}}\right)-q_\alpha\frac{\pi^2\textrm{Pm}^3}{a^4 \textrm{Pr}^2}\textrm{Ro}\sqrt{\textrm{Ta}}\right)\right], \]
\[ \delta=\frac{\pi a^2\sqrt{\textrm{Ta}}}{a^4+\pi^2\textrm{Q}-k_c^2q_\alpha R_\alpha}\cdot\left[(1+\textrm{Ro})(a^4-k_c^2q_\alpha R_\alpha)+\pi^2\textrm{QPm}(\textrm{PmRo}-1)\right]-\]
\[ -\frac{\pi a^2k_c^2R_\alpha \widetilde{\textrm{Q}}(1+\textrm{Pm})}{a^4+\pi^2\textrm{Q}-q_\alpha R_\alpha k_c^2}+
\frac{\pi^3\textrm{Pm}^2}{a^2}\textrm{QRo}\sqrt{\textrm{Ta}},\]
\[ \mathscr A_2=\frac{k_c^2\textrm{Ra}_2}{a^2\Pr}(1+q_\alpha\Pi_\alpha),\]
\[ \mathscr A_3=\frac{k_c^4\textrm{Ra}_c}{8a^4\Pr}\cdot\frac{(1+q_\alpha\Pi_\alpha)(1-q_\alpha^{(2)}\Pi_\alpha)^2(a^4+\pi^2\textrm{Q})}{a^4+\pi^2\textrm{Q}-k_c^2q_\alpha R_\alpha}+\]
\[+R_\alpha\cdot\frac{\pi^2k_c^4\textrm{Q}\textrm{Pm}^{-1}\sqrt{\textrm{Ta}}(1+q_\alpha\Pi_\alpha)(1-q_\alpha^{(2)}\Pi_\alpha)^2}{8a^4(a^4+\pi^2\textrm{Q}-k_c^2q_\alpha R_\alpha)}.\]

\end{document}